\documentclass[review,12pt]{elsarticle}
\usepackage{lineno,hyperref}
\modulolinenumbers[5]
\journal{Computers \& Fluids}
\usepackage[normalem]{ulem}
\usepackage{color}
\usepackage{xcolor}
\usepackage{subcaption}
\usepackage{booktabs}
\usepackage{amsmath}
\usepackage{amssymb} 
\usepackage{graphicx}
\usepackage{dcolumn}
\usepackage{bm}
\usepackage{mathtools}
\usepackage{esvect}
\begin{document}
\begin{frontmatter}
\title{Turbulent bubbly channel flows: Effects of soluble surfactant and viscoelasticity}
\address[add1]{Department of Mechanical Engineering, Koc University, Istanbul, Turkey.}
\address[add2]{Linn\'{e} Flow Centre and SeRC, KTH Mechanics, Stockholm, Sweden}
\address[add3]{Faculty of Industrial Engineering, Mechanical Engineering and Computer Science, University of Iceland, Hjardarhagi 2-6, 107 Reykjavik, Iceland}
\address[add4]{Aalto University, Department of Mechanical Engineering, FI-00076 Aalto, Finland}
\cortext[mycorrespondingauthor]{Corresponding Author}
\ead{mmuradoglu@ku.edu.tr}
\author[add1]{Zaheer Ahmed}
\author[add2,add4]{Daulet Izbassarov}
\author[add2,add3]{Pedro Costa}
\author[add1]{Metin Muradoglu\corref{mycorrespondingauthor}}
\author[add2]{Outi Tammisola}
\begin{abstract}
Interface-resolved direct numerical simulations are performed to examine the combined effects of soluble surfactant and viscoelasticity on the structure of a bubbly turbulent channel flow. The incompressible flow equations are solved fully coupled with the FENE-P viscoelastic model and the equations governing interfacial and bulk surfactant concentrations. The latter coupling is achieved through a non-linear equation of state which relates the surface tension to the surfactant concentration at the interface. The two-fluid Navier-Stokes equations are solved using a front-tracking method, augmented with a very efficient FFT-based pressure projection method that allows for massively parallel simulations of turbulent flows. It is found that, for the surfactant-free case, bubbles move toward the wall due to inertial lift force, resulting in formation of wall layers and a significant decrease in the flow rate. Conversely, a high-enough concentration of surfactant changes the direction of lateral migration of bubbles, i.e., the contaminated bubbles move toward the core region and spread out across the channel. When viscoelasticity is considered, viscoelastic stresses counteract the Marangoni stresses, promoting formation of bubbly wall-layers and consequently strong decrease in the flow rate. The formation of bubble wall-layers for combined case depends on the interplay of the inertial and elastic, and Marangoni forces.
\end{abstract}
\end{frontmatter}
\section{Introduction}
Complex fluids containing polymer additives exhibit viscoelastic behavior and surfactant is often added to manipulate multiphase flows. The viscoelasticity and surfactant coexist in many multiphase flow systems of practical interest such as household products, the oil and gas industry, and heating and cooling processes. In many cases the flow can be turbulent, exhibiting chaotic and multi-scale dynamics, and thus making the flow extremely complex. Polymer additives exhibit various exotic behaviors that can be exploited to perform useful functions. For instance, addition of a very small volume fraction of polymer particles in a suspending fluid can lead up to $80\%$ drag reduction in turbulent wall-bounded transport~\cite{White}. This phenomenon is very sophisticated and its full understanding remains elusive, see e.g. \cite{White}. Even the flow instabilities leading to the onset of turbulence are greatly modified by the presence of polymers \cite{Biancofore}. Introducing a second gaseous phase with surfactant contamination into this problem results in even more complex and exciting dynamics, as the two-fluid interface dynamics is tightly coupled with the elastic turbulence and the local surfactant concentration.

Numerous studies have analysed the effect of solid particles and drops or bubbles under gravity in an otherwise quiescent non-Newtonian fluid, see e.g., \citet{Zenit} for a recent review. Other studies have focused on the rheology of such suspensions in highly laminar low Reynolds number shear flows~\cite{D'Avino15}. Cross-stream migration of particles is of particular interest as it plays a critical role in many engineering and biological flows~\cite{Leal-80-arfm}.  Studies on the effect of viscoelasticity on the lateral migration of a droplet suggest various opposing scenarios of migration towards wall or center of a channel. \citet{Chan-L-79-jfm} performed an analysis based on the small perturbations using a second order fluid model and found that the viscoelasticity contained in either phase promotes migration away from the wall. \citet{mukherjee2013effects} examined the motion of a Newtonian droplet in a viscoelastic medium using a front-tracking method at a low Reynolds number and proposed a semi-analytic theory to explain the observed behavior based on the numerical simulations. In contrast with findings of \citet{Chan-L-79-jfm}, they found that there is a net viscoelastic lift force acting on the droplet stemming from the normal stress differences, which drives a droplet towards the wall. More recently, \citet{hazra2019lateral} experimentally examined the cross-stream motion of a viscoelastic drop in a viscoelastic medium in the Stokes flow regime. They proposed a presence of a viscoelasticity-induced lift force in the droplet-phase driving a droplet towards center of a channel. \citet{d2017particle} and \citet{yuan2018recent} have provided detailed reviews on the migration of a particle in a viscoelastic fluid. The detailed numerical simulations by~\citet{mukherjee2013effects} provide an explanation for the conflicting results about the cross-stream migration of a droplet in a viscoelastic continuous phase. They pointed out the importance of drop inclination (i.e., its alignment with the flow) as well as drop deformation. They showed that viscoelasticity reduces drop inclination and deformation varies non-monotonically, and the subtle competition between these two effects determines the net effect of matrix viscoelasticity on cross-stream migration of droplet. Their  careful numerical and theoretical analysis reveals that the matrix viscoelasticity acts to drive the droplet towards the channel wall due to larger effect of non-Newtonian contribution to the difference of first and second normal non-Newtonian stresses. Their study clearly demonstrates importance of detailed numerical simulations for understanding of subtle effects of viscoelasticity on multiphase flows even in the absence of inertial at very low Reynolds numbers. Strong non-linear interactions of viscoelasticity and inertia make the problem much harder and severely limit predictive capability of perturbative analysis, and thus require detailed direct numerical simulations. 
 
Direct numerical simulations (DNS) have been performed to examine behavior of bubbly channel flows for both cases of nearly spherical and deformable bubbles in the absence of viscoelasticity and surfactant. In the case of nearly spherical bubbles, \citet{Lu06} found that the clean (i.e., surfactant-free) bubbles tend to form a layer at the wall for upflow, while bubble-free wall layers are formed for downflow in vertical laminar flows. \citet{Lu08} have also examined the bubbly flows in the turbulent regime using a similar computational setup. They showed that, for nearly spherical bubbles, the formation of wall layers results in a reduction of the liquid flow rate, and the layered bubbles form horizontal clusters, whereas more deformable bubbles, instead, move toward the channel center. Later, \citet{Lu13} performed simulations for a larger system containing over hundred bubbles at a larger Reynolds number. The overall structure of the bubbly flow has been found to be similar to that of the smaller system. They also observed the formation of bubble clusters similar to the experimental observations of \citet{Takagi} for weakly-contaminated bubbly flows. Numerous other studies of DNS of turbulent bubbly flows can be found in the recent review by \citet{Elgobashi}.

Surfactant has a drastic effect on the behavior of bubbly flows \citep{Tasoglu, Olgac, Takagi}. \citet{Takagi} performed experiments to investigate the effects of different surfactants on the structure of upward turbulent bubbly flows in a vertical channel. In the surfactant-free case, they found that nearly spherical bubbles tend to form layers at the wall, where they merge during the rising process, and subsequently become deformable and move toward the channel center. When a small amount of surfactant, just enough to prevent coalescence, is added to the bulk fluid, the bubbles remain spherical, tend to move toward the wall and form bubble clusters. Conversely, with a significant surfactant concentration, the bubbles do not tend to move toward the channel wall but spread out across the channel. Motivated by this experimental study, \citet{Muradoglu} and \citet{AhmedetalIJMF} studied the effects of soluble surfactants on lateral migration of a single bubble in a pressure-driven channel flow. They both found that surfactant-induced Marangoni stresses act to move the bubble away from the wall and bubble stabilizes at the channel centerline in the presence of sufficiently strong surfactant contamination. \citet{Lu} examined the effect of insoluble surfactants on the structure of turbulent bubbly upflow in a vertical channel by DNS. They showed that surfactant contamination prevents  the formation of bubble-rich wall layers and bubbles are uniformly distributed across the channel cross-section. \citet{pesci2018computational} computationally examined the dynamics of single rising bubbles influenced by soluble surfactant. They observed that the quasi steady state of the rise velocity is reached without ad- and desorption being necessarily in equilibrium. Soligo et al.~\cite{soligo2019coalescence,soligo2019breakage} developed a modified phase field method (PFM) for simulations of turbulent flows with large and deformable surfactant-laden droplets. They have used this method to examine breakage/coalescence rates and size distribution of surfactant-laden droplets in turbulent flow. 

Even though effects of polymers and surfactant have been studied separately, their interactions and combined effects on pressure-driven turbulent bubbly channel flows remain a challenging problem that has not been explored experimentally or computationally.  Only a few numerical studies~\cite{nabiMsthesis,panigrahi2018deformation,venkatesan2019simulation} have been recently done to examine the effect of surfactant and viscoelasticity on the deformation and shape of a buoyancy-driven droplet. In this regard, the three-dimensional interface-resolved numerical simulations can provide significant insight about this highly complex phenomenon. In the present study, we report the results of the first direct numerical simulations where the combined effects of viscoelasticity and surfactant on turbulent bubbly flows are investigated. 

\section{Computational setup and numerical method}
\begin{figure}[!htb]
\centering
\includegraphics[width=0.48\textwidth,trim={3.5cm 0.2cm 3.5cm 0.2cm},clip]{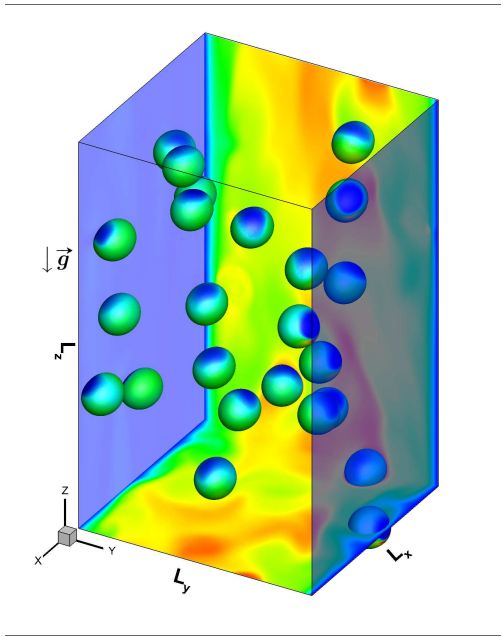}
    \caption{Schematic representation of the computational setup considered in the present work. The contours on the plane represent streamwise velocity with the scale ranging from $0.15W^*_b$ (blue) to $1.1W^*_b$ (red) where $W^*_b$ is the average bulk velocity in the single-phase case. The contours on the bubble surface represent interfacial surfactant concentration ($\Gamma$) with the scale ranging from $0$ (blue) to $0.15$ (red).}
\label{fig1}
\end{figure}

We consider vertical turbulent channel flow as shown in Fig.~\ref{fig1}. The flow is periodic in the spanwise ($x$) and streamwise ($z$) directions, while no-slip/no-penetration boundary conditions are applied at the walls ($y$ direction). The flow is driven upward by an imposed constant pressure gradient $\frac{\mathrm{d}p^*}{\mathrm{d}z^*}=constant$. Note that dimensional quantities are denoted by superscript $^*$ in the present study, e.g., $p^*$ and $p$ represent the dimensional and non-dimensional pressure, respectively. At a statistically steady state, the average wall shear stress $\tau^{*}_w$ is related to the pressure gradient and the weight of the bubble/liquid mixture by a streamwise momentum balance: 
\begin{equation}
\tau^{*}_w = - \left(\frac{\mathrm{d}p^*}{\mathrm{d}z^*}+\rho^*_{av}g^*\right)h^*=-\mathcal{B^*} h^*,
\label{tau}
\end{equation}
\noindent where $\rho^*_{av}$ is the average density in the mixture and $h^*$ is the channel half-width. Note that the value of $\mathcal{B^*}$ dictates the bulk flow direction, i.e.,\ upflow ($\mathcal{B^*}<0$) and downflow ($\mathcal{B^*}>0$). In all the cases, undeformed spherical bubbles are placed randomly in a vertical channel in an initially single-phase statistically fully developed turbulent flow, with a shear Reynolds number of $Re_\tau = w_{\tau} h/\nu_o = 127.3$. Here, the friction velocity is $w^*_\tau = \sqrt{\tau^*_w /\rho^*_o}$ , where $\rho^*_o$ and $\nu^*_o$ are the density and kinematic viscosity of the liquid respectively. Before the addition of the bubbles, the channel Reynolds number, based on the average bulk velocity and full channel width is about $Re_{bulk} \approx 3786$. The single-phase turbulent channel flow is generated by initialising the flow field with a streamwise-aligned vortex pair~\cite{Henningson}, which effectively triggers transition to turbulence. The physical parameters governing the flow are listed in Table~\ref{table:parametrs} where subscripts ``$i$" and ``$o$" denote the properties of the inner (dispersed) and the outer (continuous) fluids, respectively, and $\lambda^*$ is the polymer relaxation time. The Morton number ($M= g^*{\mu^*_o}^{4}/\rho^*_o {\sigma^*_s}^{3}$) used here is $M=6.17\times 10^{-10}$ which is higher than the Morton number of $M = 2.52\times10^{-11}$ for an air bubble in water at $20^\circ$C but could be matched by using an aqueous solution of sugar~\cite{stewart1995bubble}.

The incompressible flow equations for a viscoelastic fluid are discretized using a second-order finite-difference/front-tracking (FD/FT) method~\citep{Tryggvason}. Following~\cite{Tryggvason},~\cite{Muradoglu} and~\cite{Izbassarov15}, a single set of governing equations can be written for the entire computational domain provided the jumps in the material properties are taken into account and the effects of the interfacial surface tension are treated appropriately. The governing equations are non-dimensionalized using a length scale $\mathcal{L^*}$, a velocity scale  $\mathcal{U^*}$ and a time scale $\mathcal{T^*} =\frac{\mathcal{L^*}}{\mathcal{U^*}}$. The density and viscosity are non-dimensionalized using the density $\rho^*_o$ and total viscosity $\mu^*_{o}$ of the continuous phase while the surface tension is normalized by the surface tension of the surfactant-free surface tension $\sigma^*_s$. The non-dimensional momentum equation, accounting for interphase coupling, is than given by
\begin{eqnarray}
\frac{\partial{\rho}{\mathbf{u}}}{\partial t}+\nabla\cdot({\rho}{\mathbf{u}}{\mathbf{u}})&=&-\nabla{p}+\frac{(\rho-\rho_{av}){\mathbf{g}}}{Fr^2} + \nabla\cdot{\boldsymbol\tau} 
+\frac{1}{Re}\nabla\cdot\mu_s(\nabla{{\mathbf{u}}}+\nabla{{\mathbf{u}}^T}) \nonumber \\
&+&\frac{1}{We}\int_A\big[\sigma(\Gamma)\kappa{\bf{n}}+\nabla_s\sigma(\Gamma)\big]\delta({\bf{x}}-{\bf{x_f}})dA,
\label{NS}
\end{eqnarray}
where $\rho$, ${\mathbf{u}}$,  $p$, $\pmb{\tau}$, $\mu_s$ and $\sigma$ are the non-dimensional density, the velocity vector,  the pressure, the polymer stress tensor, the solvent viscosity and the surface tension coefficient, respectively. The non-dimensional gravitational acceleration ${\mathbf{g}}$ is the unit vector pointing in the direction of the gravitational acceleration. In Eq.~(\ref{NS}), the last term on the right hand side represents the surface tension where $A$ is the surface area, $\kappa$ is twice the mean curvature, $\mathbf{n}$ is the unit normal vector and $\nabla_s$ is the gradient operator along the interface defined as
\begin{align}
    \nabla_s=\nabla-{\mathbf{n}}({\mathbf{n}}\cdot\nabla).
\end{align} 
We emphasize that the surface tension is a function of the interfacial surfactant concentration $\Gamma$ and $\nabla_s\sigma(\Gamma)$ represents the surfactant-induced Marangoni stress. The non-dimensional numbers in Eq.~(\ref{NS}) are Reynolds number $(Re=\rho^*_o\mathcal{U^*}\mathcal{L^*}/\mu^*_o)$, Froude number $(Fr=\frac{\mathcal{U^*}{}}{\sqrt{g^*\mathcal{L^*}}})$ with $g^*$ being the magnitude of the gravitational acceleration and the Weber number $(We= \frac{\rho^*_o\mathcal{U^*}{^2}\mathcal{L^*}}{\sigma^*_s})$.
\begin{table}[!htb]
\caption{Physical and computational parameters governing the flow.}
\label{table:parametrs}
\begin{tabular}{l l  l }
\hline
\hline
Friction Reynolds number    & $Re_\tau$           & $127.3$\\ 
Domain size     & $L_x \times L_y \times L_z$           & $\pi/2 \times 2 \times \pi$\\ 
Resolution      & $N_x \times N_y \times N_z$           & $128 \times 192 \times 256$\\
Inner-scaled resolution   & $\Delta x^+ =\Delta z^+ $   & 1.56\\
Inner-scaled resolution   & $\Delta y^+$ min/max        & 0.76/1.72\\ 
Channel half-width & $h^*$                                 & 1.0     \\
Bubble diameter & $d^*_b$                                 & 0.3     \\
Void fraction        &   $\Phi$                               & 3.44\%  \\
Density and viscosity ratio  & $\rho^*_i/\rho^*_o$, $\mu^*_i/\mu^*_o$    & 0.05, 0.1   \\
Total pressure gradient & $\mathcal{B^*}=\frac{dp^*}{dz^*}+\rho^*_{av}g^*$   & -0.0018 \\
Surface tension &  $\sigma^*_s$                             & 0.01    \\ 
Gravitational acceleration & $g^*$                        & 0.05    \\
Single-phase average bulk velocity & $W^*_b$                      & 0.63   \\
E$\mathrm{\ddot{o}}$tv$\mathrm{\ddot{o}}$s Number & $Eo= \rho^*_o g^*d^{*2}_{b}/\sigma^*_s $	& 0.45 \\
Morton Number & $M= g^*{\mu^*_o}^{4}/\rho^*_o {\sigma^*_s}^{3}$ 	& 6.17$\times 10^{-10}$ \\
Peclet number  & $Pe_c=Pe_s=W^*_b d^*_b/D^*_c$                      & 252 \\
Biot Number & $Bi=k^*_d d^*_b/W^*_b$                          & 0.95 \\ 
Damkohler Number & ($Da=\Gamma^*_{max}/d^*_bC^*_\infty$)           & 13.3/6.6/3.3 \\ 
Langmuir number &($La=k^*_a C^*_\infty/k^*_d$)                     & 0.0625/0.125/0.25 \\
Elasticity Number & $\beta_s=R^*T^*\Gamma^*_\infty/\sigma^*_s$    & 0.5\\ 
Weissenberg Number & $Wi=\lambda^* W^*_b/2h^*$                & 0/5/10  \\
Solvent viscosity ratio & $\beta=\mu^*_s/\mu^*_o$               & 0.9 \\
Extensibility parameter & $L$                               & 60 \\
\hline
\hline
\end{tabular}
\end{table} 

The fluid properties remain constant along material lines, i.e.,
\begin{align}
\frac{D\rho}{Dt} &=0, & \frac{D\mu_s}{Dt} &=0, & \frac{D\mu_p}{Dt} &=0, & \frac{D\lambda}{Dt} &=0,
\end{align}
where $\frac{D}{Dt}=\frac{\partial{}}{\partial{t}}+\mathbf{u}\cdot\nabla$ is the material derivative. The indicator function is used to set the material properties in each phase and is defined as 
\begin{align}
I({\mathbf{x}},t) &=
\begin{cases}
 1    & \text{in bulk fluid,}\\
 0    & \text{in bubble domain.}
\end{cases}
\end{align}
Then a material property, $\phi$, is specified as
\begin{align}
\phi =\phi_{o}I({\mathbf{x}},t)+\phi_i\left(1-I({\mathbf{x}},t)\right),
\end{align}
where the subscripts ``$i$" and ``$o$" denote the properties of the bubble and bulk fluid, respectively. 

The present soluble surfactant methodology is same as that developed by\citet{muradoglu2008front,Muradoglu}, and is described below for completeness. The $C_\infty~(=\frac{C_\infty^*}{C^*_{ref}})$ represents the non-dimensional initial bulk surfactant concentration  and $\Gamma~(=\frac{\Gamma^*}{\Gamma^*_{max}})$ represents the non-dimensional interfacial surfactant concentration, where $C^*_{ref}$ and $\Gamma^*_{max}$ denote the reference bulk surfactant concentration taken as the critical micelle concentration (CMC) and the maximum packing concentration, respectievely. Surface tension coefficient is scaled by $\sigma^*_s$. The local value of the surface tension coefficient is related to the local surfactant concentration through the Langmuir equation of state~\cite{Levich}
\begin{align}
\sigma= 1+\beta_s\ln\left(1-\Gamma\right),
\end{align}
where $\beta_s$ is the elasticity number. Equation~\eqref{beta} is bounded to prevent nonphysical (negative) values of the surface tension:
\begin{align}
\sigma=\max\left(\epsilon_\sigma,1+\beta_s\ln\left(1-\Gamma\right)\right),
\label{beta}
\end{align}
where $\epsilon_{\sigma}$ is taken as 0.05 in the present study. Note that $\epsilon_{\sigma}$ is typically larger than this value in the common physical systems (e.g., $\epsilon_{\sigma}\sim0.2-0.3$) but it does not have any influence on the present results since the interfacial surfactant concentration remains much lower than the maximum packing concentration in all the cases considered here. The  evolution  equation for the interfacial surfactant concentration was  derived by \citet{Stone} and can be expressed in the non-dimensional form as 
\begin{align}
\frac{1}{A}\frac{D \Gamma A}{Dt}= \frac{1}{Pe_{s}}\nabla_{s}^2\Gamma+Bi\dot{S}_{\Gamma},
\label{Us}
\end{align}
where $A$ is the surface area of an element of the interface and $Pe_s=\frac{\mathcal{U^*}\mathcal{L^*}}{D^*_s}$ is the interfacial Peclet number with $D^*_s$ being the diffusion coefficient along the interface. The Biot number is defined as $Bi=\frac{k^*_d\mathcal{L^*}}{\mathcal{U^*}}$, where $k^*_d$ is the desorption coefficient. The non-dimensional source term $\dot{S}_\Gamma$ is given by
\begin{align}
\dot{S}_\Gamma=La C_s(1-\Gamma)-\Gamma,
\label{dotS}
\end{align}
where $C_s$ is the bulk surfactant concentration near the interface and $La$ is the Langmiur number defined as $La=\frac{k^*_a C^*_{ref}}{k^*_d}$ with $k^*_a$  being the adsorption coefficient. The bulk surfactant concentration is governed by an advection-diffusion equation of the form
\begin{align}
\frac{\partial C}{\partial t}+\nabla\cdot(C{\mathbf{u}})=\frac{1}{Pe_c}\nabla\cdot(D_{co}\nabla C),
\label{bulkconc}
\end{align}
where $Pe_c =\frac{\mathcal{U^*}\mathcal{L^*}}{D^*_{c}}$ is the Peclet number based on bulk surfactant diffusivity. The coefficient $D^*_{co}$ is related to the molecular diffusion coefficient $D^*_c$ and the phase indicator function $I$ as
\begin{align}
D^*_{co}=D^*_c I({\mathbf{x}},t).
\end{align}
The source term in Eq. \eqref{dotS} is related to the bulk concentration by
\begin{align}
\dot{S}_\Gamma=-\frac{1}{Pe_c Da}({\mathbf{n}}\cdot\nabla C_\mathrm{interface}).
\label{sourceterm}
\end{align}
where $Da=\frac{\Gamma^*_{max}}{\mathcal{L^*}C^*_{ref}}$ is the Damk\"{o}hler number. Following \citet{Muradoglu}, the boundary condition at the interface given by Eq.~\eqref{sourceterm} is first converted into a source term for the bulk surfactant evolution equation. In this approach it is assumed that all the mass transfer between the interface and bulk takes place in a thin adsorption layer adjacent to the interface. Thus, the total amount of mass adsorbed on the interface is distributed over the adsorption layer, and added to the bulk concentration evolution equation as a negative source term. Equation~\eqref{bulkconc} thus takes the following form:
\begin{align}
\frac{\partial C}{\partial t}+\nabla\cdot(C{\mathbf{u}})=\frac{1}{Pe_c}\nabla\cdot(D_{co} \nabla C)+\dot{S}_c,
\label{BulkSurf}
\end{align}
where $\dot{S}_c$ is the source term evaluated at the interface and distributed onto the adsorption layer in a conservative manner. The details of this treatment can be found in \cite{Muradoglu}.

The FENE-P model is adopted as the constitutive equation for the polymeric stress tensor $\pmb{\tau}$. This constitutive equation and its numerical solution are described in details in \citep{Izbassarov15,Izbassarov18}. The model can be written as
\begin{equation}
 \label{stress_eq}
\begin{split}
\frac{\partial \textbf{B}}{\partial t} + \nabla\cdot(\textbf{u}\textbf{B}) 
- (\nabla\textbf{u})^T\cdot\textbf{B} - \textbf{B}\cdot\nabla\textbf{u} = 
-\frac{1}{Wi}\left(F\textbf{B}-\textbf{I}\right), \\
F=\frac{L^2}{L^2-\mathrm{trace}(\textbf{B})},
\end{split}
\end{equation}
\noindent where $\textbf{B}$, $Wi$, $L$, and $\textbf{I}$ are the conformation tensor, the Weissenberg number defined as $Wi = \frac{\lambda^* \mathcal{U^*}}{\mathcal{L^*}}$ with $\lambda^*$ being the polymer relaxation time, the maximum polymer extensibility, and the identity tensor, respectively. Once the conformation tensor is obtained from Eq.~(\ref{stress_eq}), the polymeric stress tensor is computed as
\begin{eqnarray}
\pmb{\tau} = \frac{1}{Re Wi}(1-\beta)\left(F\textbf{B} - \textbf{I}\right),
\label{ConfTensor}
\end{eqnarray}
\noindent where $\beta=\mu^*_s/\mu^*_o$ is the solvent viscosity ratio. 

The viscoelastic constitutive equations are highly non-linear and become extremely stiff at high Weissenberg numbers.  Thus, their numerical solution is generally a formidable task and is notoriously difficult especially at high Weissenberg numbers \citep{Izbassarov15, Izbassarov18}. To overcome the so-called high Weissenberg number problem, the log-conformation method is employed \citep{Fattal} in the present simulations. In this approach, Eq.~(\ref{stress_eq}) is rewritten in terms of the logarithm of the conformation tensor through eigendecomposition, i.e.,  $\pmb\Psi=\log\textbf{B}$, which ensures the positive definiteness of the conformation tensor. The core feature of the formulation is the decomposition of the gradient of the divergence free velocity field $\nabla \textbf{u}^T$ into two anti-symmetric tensors denoted by $\pmb\Omega$ (pure rotation) and $\textbf{N}$, and a symmetric tensor denoted by $\textbf{D}$ which commutes with the conformation tensor~\cite{Fattal}, i.e.,
\begin{eqnarray}
 \nabla \textbf{u}^T = \pmb{\Omega} + \textbf{D} + \textbf{NB}^{-1}.
 \label{gradu}
\end{eqnarray}
Inserting Eq.~(\ref{gradu}) into Eq.~(\ref{stress_eq}) and replacing the conformation tensor with the new variable $\pmb\Psi$, the transformed constitutive equations can be written as
\begin{eqnarray}
\frac{\partial \pmb\Psi}{\partial t} + \nabla\cdot(\textbf{u}\pmb\Psi) 
- (\pmb\Omega \pmb\Psi - \pmb\Psi \pmb\Omega) - 2\textbf{D}
&=& \frac{F}{Wi}(e^{-\pmb\Psi} - \textbf{I}).
\label{TransformedConTensor}
\end{eqnarray}
Once Eq.~(\ref{TransformedConTensor}) is solved for $\pmb\Psi$, the conformation tensor is then obtained using the inverse transformation as $\textbf{B} = e^{{\pmb\Psi}}$. 

\subsection{Flow Solver}
The flow equations (Eq.~(\ref{NS})) are solved fully coupled with the FENE-P model (Eq.~(\ref{TransformedConTensor})) and the interfacial and bulk surfactant concentration evolution equations (Eqs.~(\ref{Us}) and (\ref{BulkSurf})) using a front-tracking/finite-difference (FD/FT) method \citep{Tryggvason,Muradoglu,Izbassarov15}.  All the field equations are solved on a fixed Eulerian grid with a staggered arrangement where the velocity nodes are located at the cell faces while the material properties, the pressure, the bulk surfactant concentration and the extra stresses are all located at the cell centers\citep{Izbassarov15}. The interfacial surfactant concentration evolution equation is solved on a separate Lagrangian grid \citep{Muradoglu}. The spatial derivatives are discretized with second-order central differences for the diffusive terms, while the convective terms  are discretized using a QUICK scheme~\citep{Leonard} in the momentum equation, and a fifth-order WENO-Z \citep{Borges} scheme in the viscoelastic and the bulk surfactant concentration equations. The equations are integrated in time with a second-order predictor-corrector method in which the first-order solution (Euler method) serves as a predictor that is then corrected by the trapezoidal rule \citep{Tryggvason}.  

    The details of the front-tracking method can be found in the review paper by \citet{Tryggvason} and in the recent book by \citet{Tryggvason-book}. The numerical methods for the treatment of soluble surfactant and viscoelasticity have been fully discussed by \citet{Muradoglu} and \citet{Izbassarov15}, respectively. The new ingredient of the numerical method in the present study is the FFT-based pressure solver which is briefly described here. Since the projection method requires solution of a variable-density Poisson equation in the multiphase flows, FFT-based solvers cannot be  used directly. To overcome this difficulty, the pressure-splitting technique presented in \cite{Dong} and \cite{Dodd} is adopted, allowing for a direct and fast solution of a constant-coefficients Poisson equation using the same FFT-based solver.  Regarding the FFT-based pressure solver, the overall procedure to advance time from time step $n$ to the first-order predicted solution at time step $n+1$ is provided below:
\begin{eqnarray}
 \frac{\rho^{n+1} \textbf{u}^{\star} -\rho^n\textbf{u}^n}{\Delta t}
  =   \left[-\nabla_h \cdot (\rho \textbf{uu}) 
 +\frac{1}{Re}\nabla_h\cdot(\mu_{s}(\nabla_h\textbf{u}+\nabla_h^T\textbf{u}))
 + \frac{(\rho-\rho_{av}){\mathbf{g}}}{Fr^2}  \right. \nonumber\\
 + \nabla \cdot \pmb{\tau}+ \frac{1}{We}\int_A \big[\sigma(\Gamma)\kappa{\bf{n}}+\nabla_s\sigma(\Gamma)\big]\delta({\bf{x}}-{\bf{x_f}})dA \left. \right]^n,   \label{projectionstep}
\end{eqnarray}
\begin{eqnarray}
 \nabla^2 p^{n+1} =  \nabla \cdot \left[\left(1-\frac{\rho_0}{\rho^{n+1}}\right)\nabla \left(2 p^{n}-p^{n-1}\right) \right]  + \frac{\rho_0}{\Delta t}\nabla \cdot {\bf u}^{\star}, 
 \label{poisson LS}
\end{eqnarray}
 \begin{eqnarray}
{\bf u}^{n+1} = {\bf u}^{\star} -\Delta t\left[\frac{1}{\rho_0} \nabla p^{n+1}+\left(\frac{1}{\rho^{n+1}}-\frac{1}{\rho_0}\right) \nabla \left(2 p^{n}-p^{n-1}\right) \right],
\end{eqnarray}
 \begin{eqnarray}
  \pmb\Psi^{n+1} = \pmb\Psi^n + \Delta t \left(-\nabla\cdot(\textbf{u}\pmb\Psi) 
 + (\pmb\Omega \pmb\Psi - \pmb\Psi \pmb\Omega) + 2\textbf{D}
 + \frac{F}{Wi}(e^{-\pmb\Psi} - \textbf{I})\right)^n,
 \label{projection LS}
 \end{eqnarray}
where $\nabla_h$ is the discrete version of the nabla operator, $\rho_0 = \min(\rho_i,\rho_o)$ and $\textbf{u}^{\star}$ is the unprojected velocity. The solution at this stage provides an estimate at the new time level and is used to compute the solution at the time level $n+2$. Then solutions at time levels $n$ and $n+2$ are averaged to obtain a second order solution at the new time level $n+1$ as described by \citet{Tryggvason}. Note that the overall time-integration scheme is equivalent to the trapezoidal rule for the linear problem. 
\subsection{Parallelization}
The parallelization method developed by \citet{Farooqi} is slightly modified for the FFT-based pressure solver in the present study. First, the previous iterative multigrid Poisson solver in the HYPRE library \cite{HYPRE-lib}, the most demanding part of the numerical method, has been replaced with a versatile and efficient FFT-based direct solver for a constant-coefficients Poisson equation of the DNS code \emph{CaNS}; see \cite{Costa} for more details. The extension of the solver for the phase indicator function with this approach is trivial. To determine the overall speed up of the numerical algorithm, simulations are performed for a fully developed laminar flow using the  domain size of $2\times 2 \times 2$ discretized by $512^3$ and 64 randomly distributed bubbles. Number of cores for the domain and fronts are 256 and 64, respectively. The results are summarized in Table~\ref{table:solver}. The FFT solver satisfies the divergence-free constraint on the velocity field to near the machine precision at every grid point (i.e., $|\nabla_h\cdot \mathbf{u}|_{max} < 10^{-14}$) whereas, the HYPRE (PFMG) reduces the residuals below a prespecified tolerance value i.e., $|\nabla_h\cdot \mathbf{u}|_{max} < \mathcal{E}_{tol}$. As can be seen, the present numerical algorithm with the FFT solver gets 2 and 8 times speed up for the error tolerances  $\mathcal{E}_{tol}= 10^{-5}$ and $\mathcal{E}_{tol}= 10^{-10}$, respectively. Note that  $\mathcal{E}_{tol}= 10^{-5}$ is sufficient to achieve a desired overall numerical accuracy for the kind of simulations presented here.

\begin{table}
\centering
\caption{Wall clock time per time step in seconds, and corresponding speedup of the DNS solver when a FFT-based Poisson solver is used instead of the iterative solver of HYPRE (PFMG).}
\vspace{-0.25cm}
\label{table:solver}
\begin{tabular}{|c |c  | c| c|}
\hline
Multigrid tolerance & HYPRE & FFT &Speedup \\
\hline
$|\nabla_h\cdot \mathbf{u}|_{max} < 10^{-5}$  & 4.7   & 2.1   & 2.2\\
$|\nabla_h\cdot \mathbf{u}|_{max} < 10^{-10}$ & 16.3  & 2.1   & 7.8\\
\hline
\end{tabular}
\end{table}

\begin{figure}[!htb]
\centering
\begin{subfigure}{0.4\textwidth}
\centering
\includegraphics[width=\textwidth, trim={0cm 0.1cm 0.5cm 0},clip]{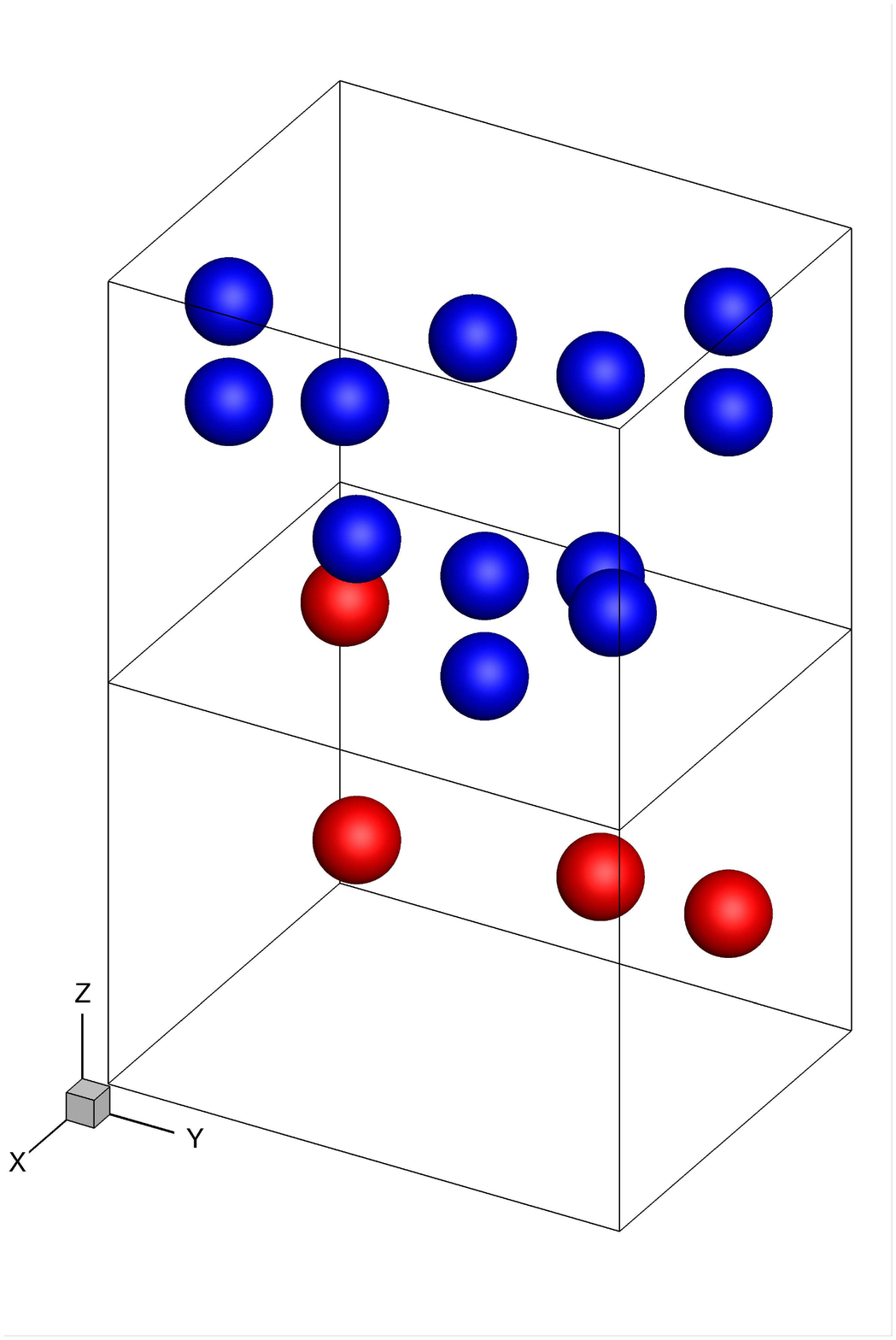}
\end{subfigure}
\begin{subfigure}{0.43\textwidth}
\centering  
\includegraphics[width=\textwidth, trim={0cm 0.1cm 1cm 0},clip]{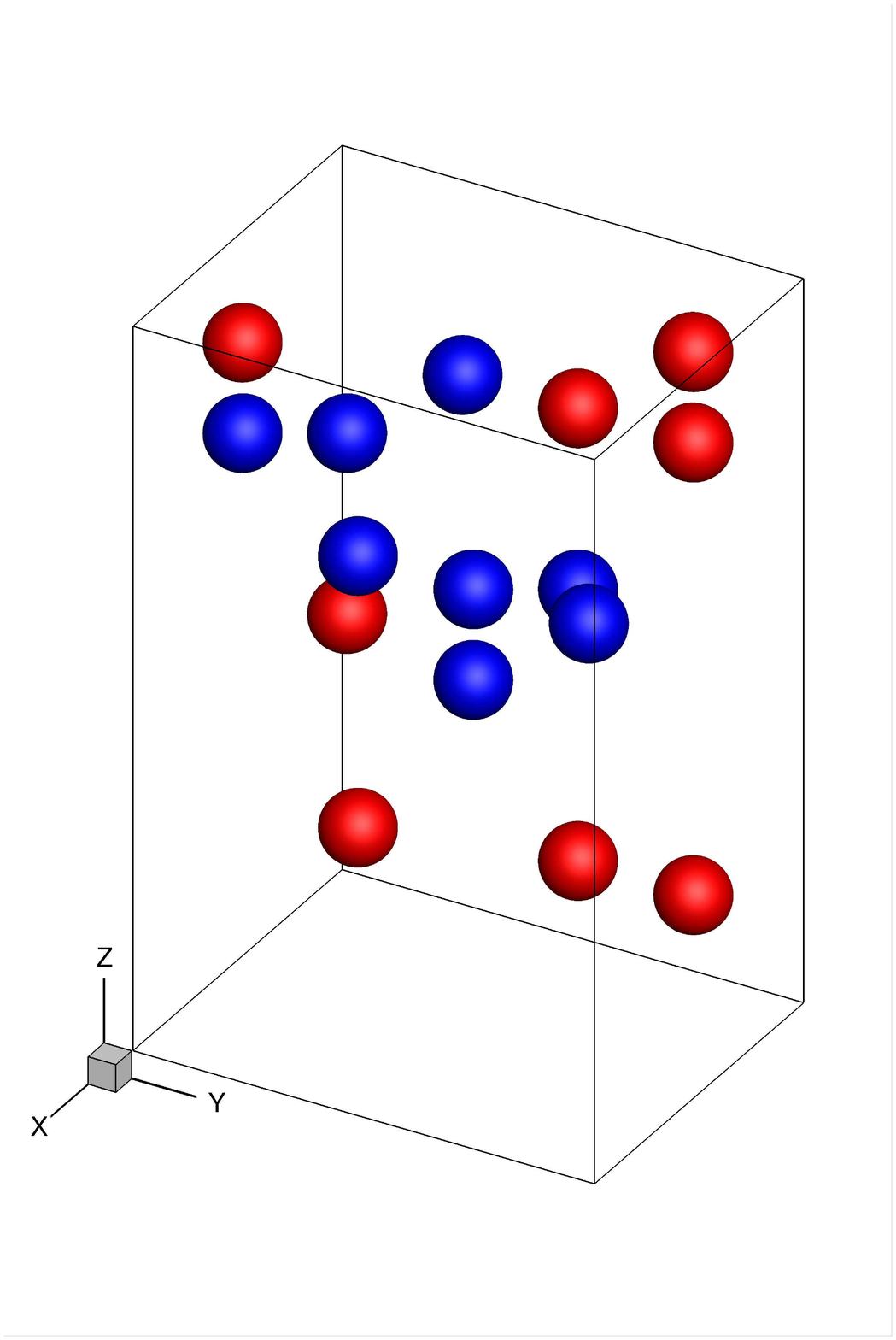}  
\end{subfigure}
\caption{(Left) Computational load distribution under Redundantly-All-Compute (RAC) scheme, in contrast to that of the Particle-Per-Front (PPF) scheme (right) where the load distribution is decoupled from the domain decomposition. Blue and red colors denote load on Rank 1 and Rank 2, respectively.}
\label{Parallelscheme}
\end{figure}

Next, the new parallelization strategy for the Lagrangian grid is implemented for the present computational setup, namely Particles-Per-Front (PPF), where \emph{Fronts} denote to the processes computing on the Lagrangian grid. The PPF is compared with our previous strategy, namely Redundantly-All-Compute (RAC)~\cite{Farooqi}. The Redundantly-All-Compute algorithm is based on domain decomposition method similar to the Eulerian grid parallelization, where each subdomain can be assigned to one MPI-process and bubbles are distributed among the parallel \textit{Fronts}. The \textit{Front} containing the center of a bubble becomes the \emph{owner} of that bubble. An example where 16 randomly distributed bubbles are assigned to 2 \textit{Fronts} is shown in the left panel of Fig.~\ref{Parallelscheme}. As can be seen, the load on 2 \textit{Fronts} is different due to the location of bubbles in the physical domain and this can continuously happen as bubbles move in the physical domain. To prevent this scenario, we parallelize the Lagrangian grid in a way that bubbles are assigned directly to the \textit{Fronts} independent of their location in the physical domain. In this approach, the load on each \textit{Front} will remain constant throughout the simulation. A typical example of this strategy where 16 randomly distributed bubbles are assigned to 2 \textit{Fronts} (8 per \textit{Front}) is shown in the right panel of Fig.~\ref{Parallelscheme}.

The parallelization of 3D Front Tracking method leads to three types of communication namely (1) \textit{Domain} to \textit{Domain}, (2) \textit{Front} to \textit{Front} and (3) \textit{Front} to \textit{Domain}. Particles-Per-Front (PPF) parallelization strategy does not effect the \textit{Domain} to \textit{Domain} communication therefore it remains the same as described by \citet{Farooqi}. The \textit{Front} to \textit{Front} communication is completely eliminated for the (PPF) strategy as the ownership of the bubbles does not change with location of bubbles in the physical domain. The total message size between a \textit{Front} and all its \textit{Domains} in a single time-step to be  
\begin{eqnarray}
M_{f2d} &= M_{send}+M_{recv}\\
M_{send}= 9\times \sum_{i=1}^{n}P_i, & M_{recv}= 3\times \sum_{i=1}^{n} P_i
\end{eqnarray}
where $P_i$ is the total number of points for the ith bubble. 
Let $d_x$ , $d_y$, and $d_z$ be the geometry of the MPI processes for the \textit{Domains}. The number of messages sent and received by the \textit{Front} in a single time-step would be $2d^3$ (Assuming $d=d_x=d_y=d_z$) because all point coordinates destined to a single \textit{Domain} could be packed in a single message. If there are f \textit{Fronts} and an approximately equal number of \textit{Domains} are assigned to each \textit{Front}, then for the RAC strategy, the number of messages and message size per \textit{Front} are reduced by a factor of f ; each Front exchanges $2d^3/f$ messages with size of $M_{f2d}/f$ . However, the actual message size per \textit{Front} can vary based on how bubbles are distributed in the physical domain. For the PPF strategy the number of messages and message size per \textit{Front} will remain same. Thus, the PPF strategy works better for the modest domain and grid size and RAC works better for large scale simulations. 

\section{Results and Discussion}            
\begin{figure}[!htb]
\begin{subfigure}{0.49\textwidth}
\centering
\includegraphics[width=\textwidth]{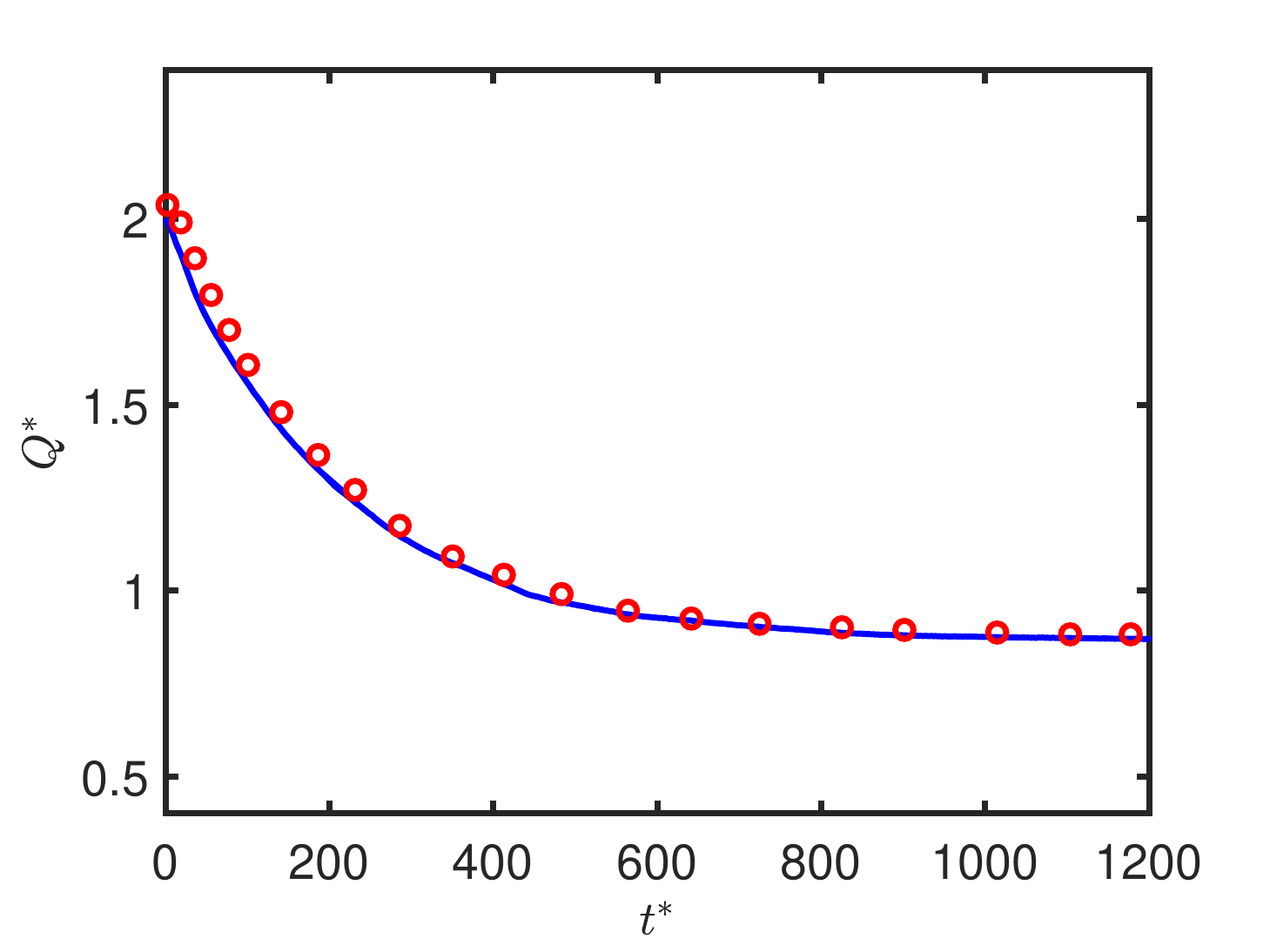}
\end{subfigure}
\begin{subfigure}{0.49\textwidth}
\centering  
\includegraphics[width=\textwidth]{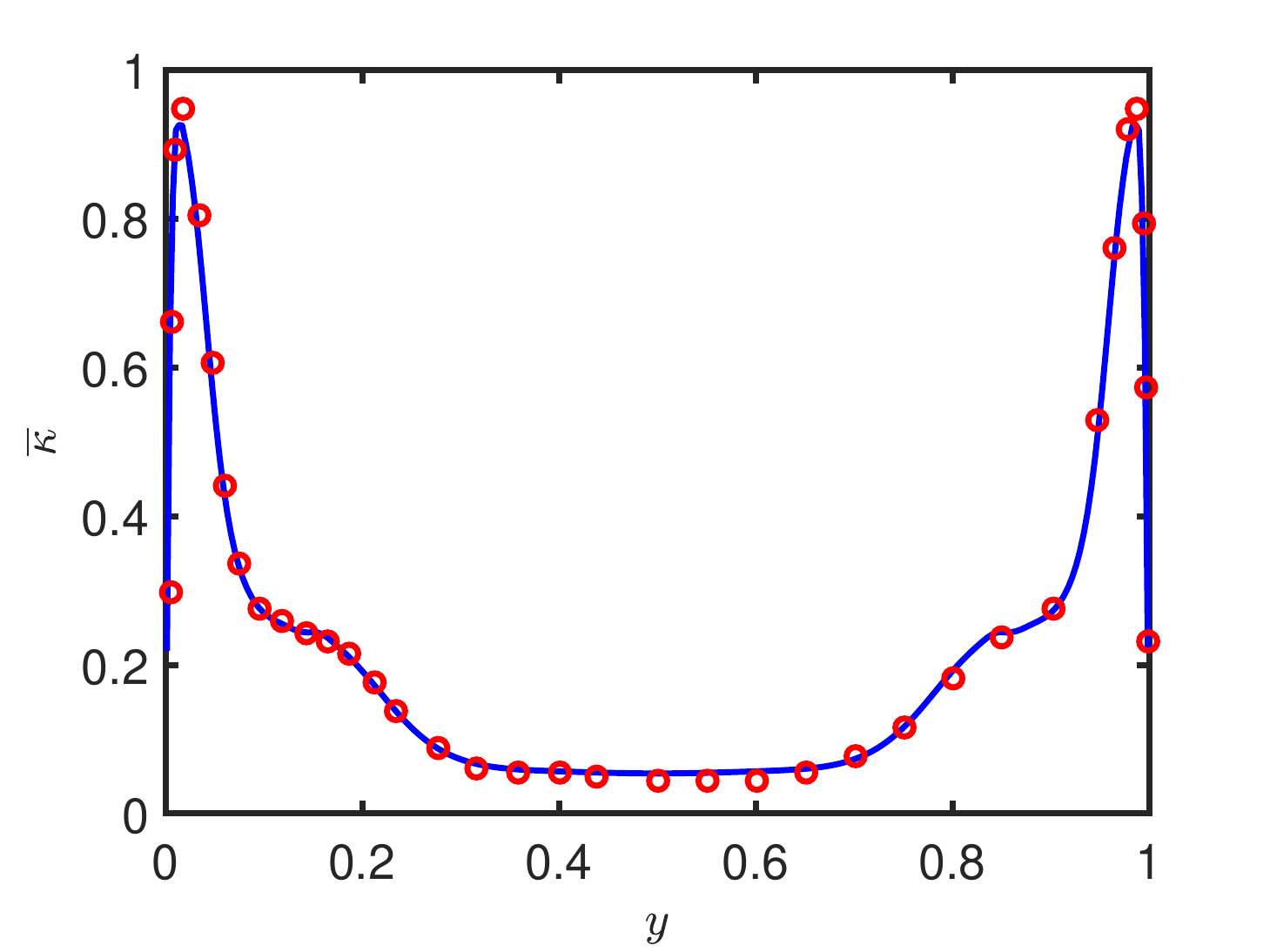}  
\end{subfigure}
\caption{The turbulent multiphase flow: (left) The evolution of the flow rate and (right) the turbulent kinetic energy at $Re_{\tau}=127.3$. The solid lines represent the current results and the symbols are the DNS data by \citet{Lu}. The turbulent kinetic energy is scaled by ${w^*_\tau}^2$}
\label{figLu}
\end{figure}

The numerical method was first validated against the DNS results of turbulent bubbly channel flow in the absence of viscoelasticity and surfactant by~\citet{Lu}. Figure~\ref{figLu} shows the time history of the liquid flow rate (panel (a)) and the profile of the mean turbulent kinetic energy (panel (b)) for the turbulent multiphase flow in a statistically steady state, compared to the DNS results by~\citet{Lu}. The present results are in excellent agreement with the results of ~\citet{Lu} showing the accuracy of the present simulations. 
The size of the rectangular domain is $\pi/2 \times 2 \times \pi$ resolved by $128\times 192\times 256$ in the spanwise ($x$), wall-normal $(y)$ and streamwise ($z$) directions respectively. This configuration have been used for various turbulent bubbly flow studies~\cite{Lu05,Lu07,Lu08,Lu} using the same numerical method i,e., finite-difference/front-tracking (FD/FT) method. This is the so called ``minimum turbulent channel'' which has been studied in detail by Jimenez and Moin~\cite{jimenez1991minimal}. The number of bubbles i.e., 24 used in the present setup is relatively a modest number. \citet{Lu13} studied the behavior of turbulent bubbly flows at much higher Reynolds number $(Re_{\tau}=250)$ and used 140 bubbles. They have also used the same domain size $\pi/2 \times 2\times \pi$ resolved by $256\times384\times512$. Thus the grid resolution used for the present setup is sufficient for the grid convergence. 

The objective of the present study is to investigate the sole and combined effects of soluble surfactant and viscoelasticity on the turbulent bubbly flows. The deformation of the bubbles has significant effect on the structure and dynamics of bubbly flows~\citep{Lu08}. To minimize the effects of deformability of the bubbles, initially, nearly spherical bubbles are included/added to the turbulent flow. The deformation of the bubbles can be characterized by two non-dimensional numbers for the present setup, i.e., the Capillary ($Ca=\mu^*_oW^*_b/\sigma^*_s$) and E$\mathrm{\ddot{o}}$tv$\mathrm{\ddot{o}}$s ($Eo= \rho^*_o g^*{d^*_{b}}^{2}/\sigma^*_s $) numbers. Both these non-dimensional numbers are kept constant and at low value, so that the bubbles remain nearly spherical for all the cases. The governing parameters related to the surfactant other than $C_\infty$ are also kept constant. Our previous study about the effects of surfactant~\citep{AhmedetalIJMF} on the dynamics of single bubble has demonstrated that the elasticity number ($\beta_s=R^*T^*\Gamma^*_\infty/\sigma^*_s$) that signifies the strength of surfactant and the bulk surfactant concentration have very similar effects. In addition, other parameters related to surfactant such as bulk and interfacial Peclet numbers, the Biot number and the Damk\"ohler number have only secondary effect on bubbly flows in statistically stationary sate~\citep{AhmedetalIJMF}. Therefore, it is deemed to be sufficient to only vary $C_\infty$ to show the effects of surfactant in the present study.  We also note that it is easy to vary $C_\infty$ by simply adding more surfactant into the bulk fluid in an experimental study. The surfactant parameters used here are based on our previous studies~\citep{AhmedetalIJMF,Muradoglu}. For the non-Newtonian flow cases, the Weissenberg number ($Wi$) quantifies the dynamic significance of the viscoelasticity. The other parameter related to the viscoelasticity is the solvent viscosity ratio ($\beta$) that essentially controls the magnitude of viscoelastic stresses as can be seen in Eq. (16). Thus only $Wi$ is varied to characterize the effects of viscoelasticity.

First, extensive simulations are performed to investigate the effects of surfactant and viscoelasticity separately. Then their combined effects are investigated. For this purpose, first computations are carried out for a range of bulk surfactant concentrations $C_\infty$ to examine the effects of surfactant on bubbly flows without viscoelasticity in section~(\ref{SrfN}). Then the simulations are repeated for three different Weissenberg numbers, quantifying the dynamic significance of the viscoelasticity, i,e., $Wi = 0$, $5$ and $10$, in the absence of surfactant to determine the sole effects of viscoelasticity in section~(\ref{Veffects}). Finally, the combined effects of surfactants and viscoelasticity are investigated in section~(\ref{SrfV}).
 
\subsection{Effects of soluble surfactant}\label{SrfN}

\begin{figure}[!htb]
\centering
\begin{subfigure}{0.49\textwidth}
\includegraphics[width=\textwidth]{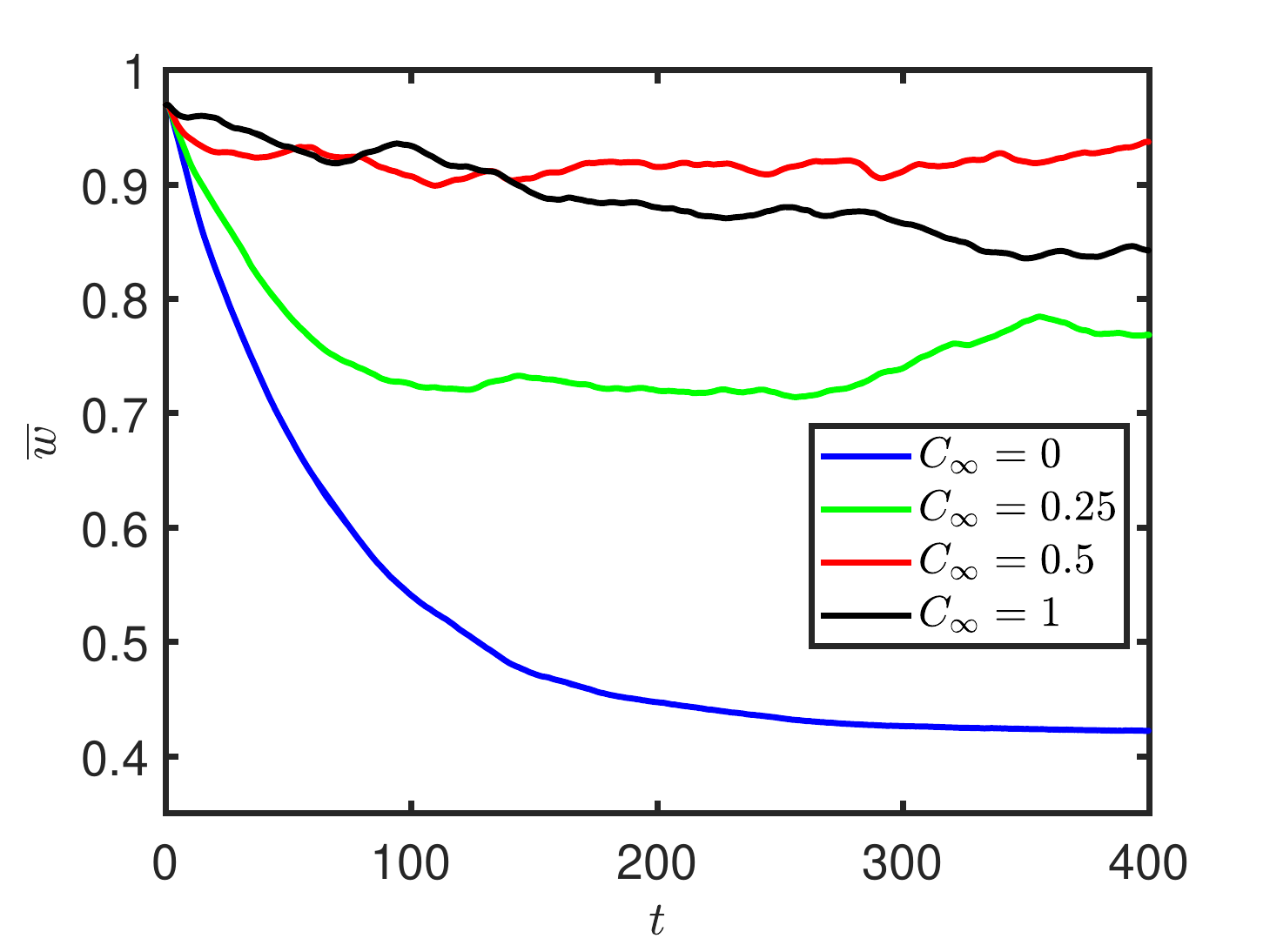}
\caption{}
\label{flow_rateN}
\end{subfigure}
\begin{subfigure}{0.49\textwidth}
\includegraphics[width=\textwidth]{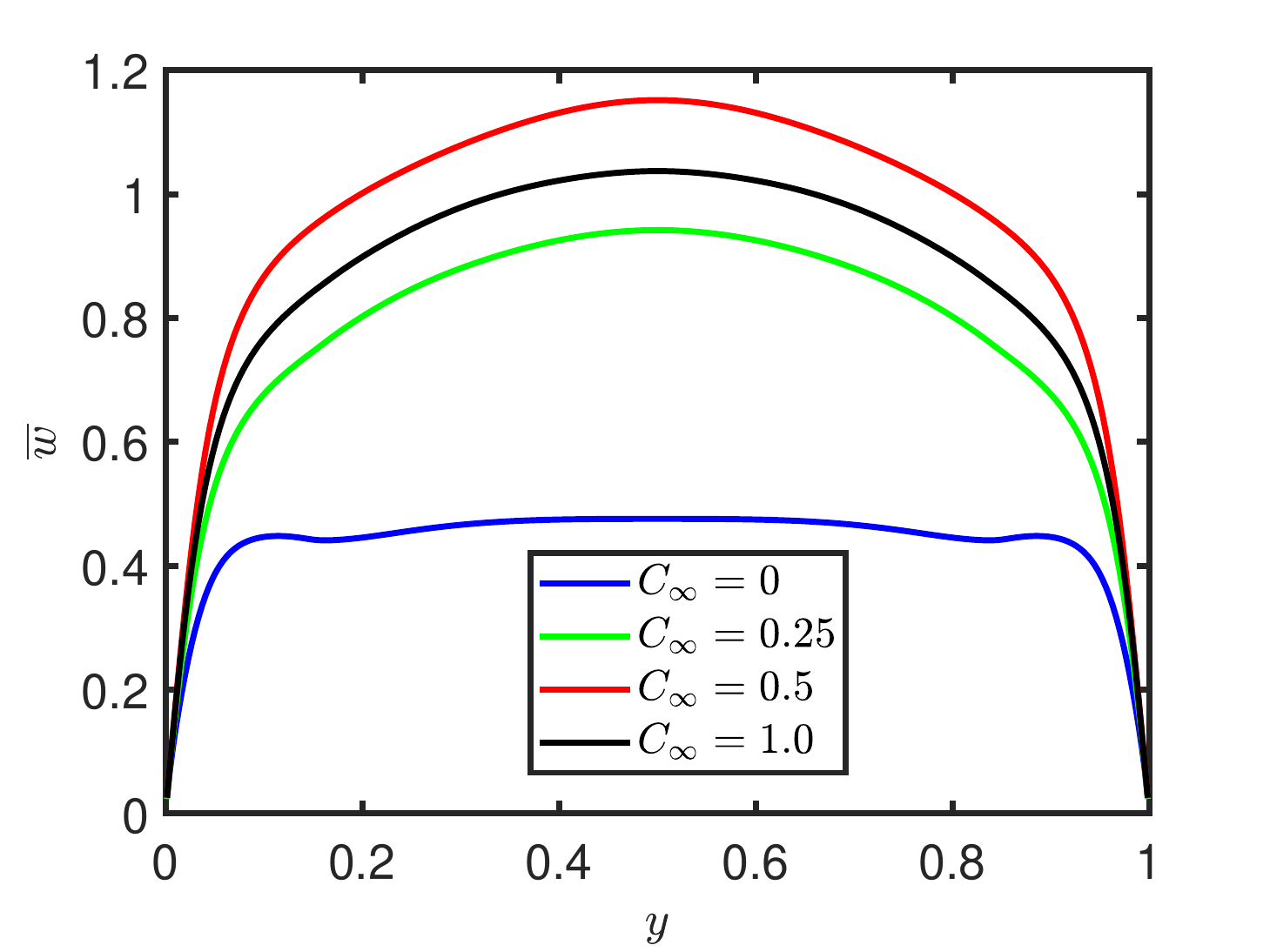}
\caption{}
\label{flow_rateN2}
\end{subfigure}
\begin{subfigure}{0.49\textwidth}
\includegraphics[width=\textwidth]{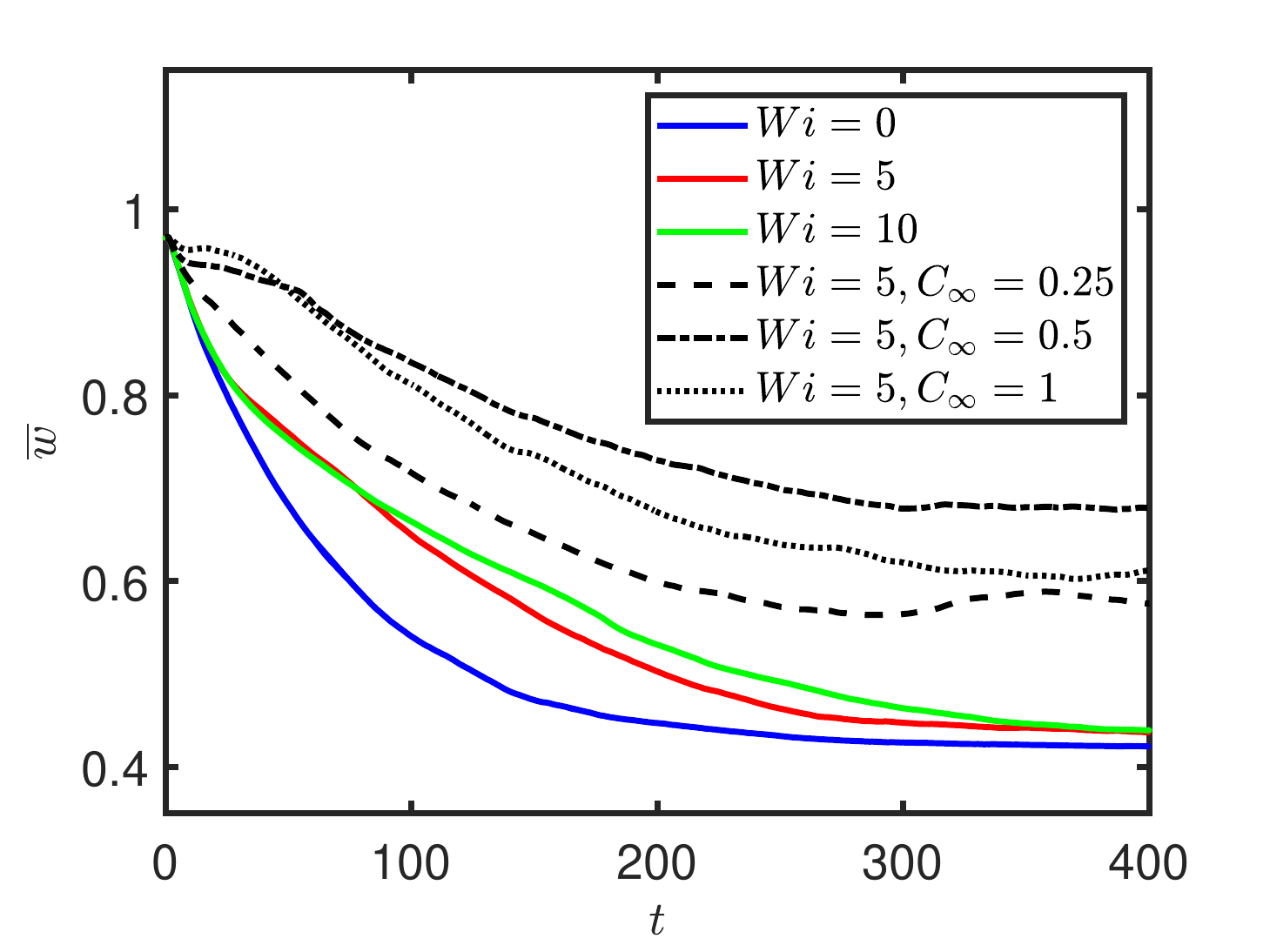}
\caption{}
\label{flow_rateV}
\end{subfigure}
\begin{subfigure}{0.49\textwidth}
\includegraphics[width=\textwidth]{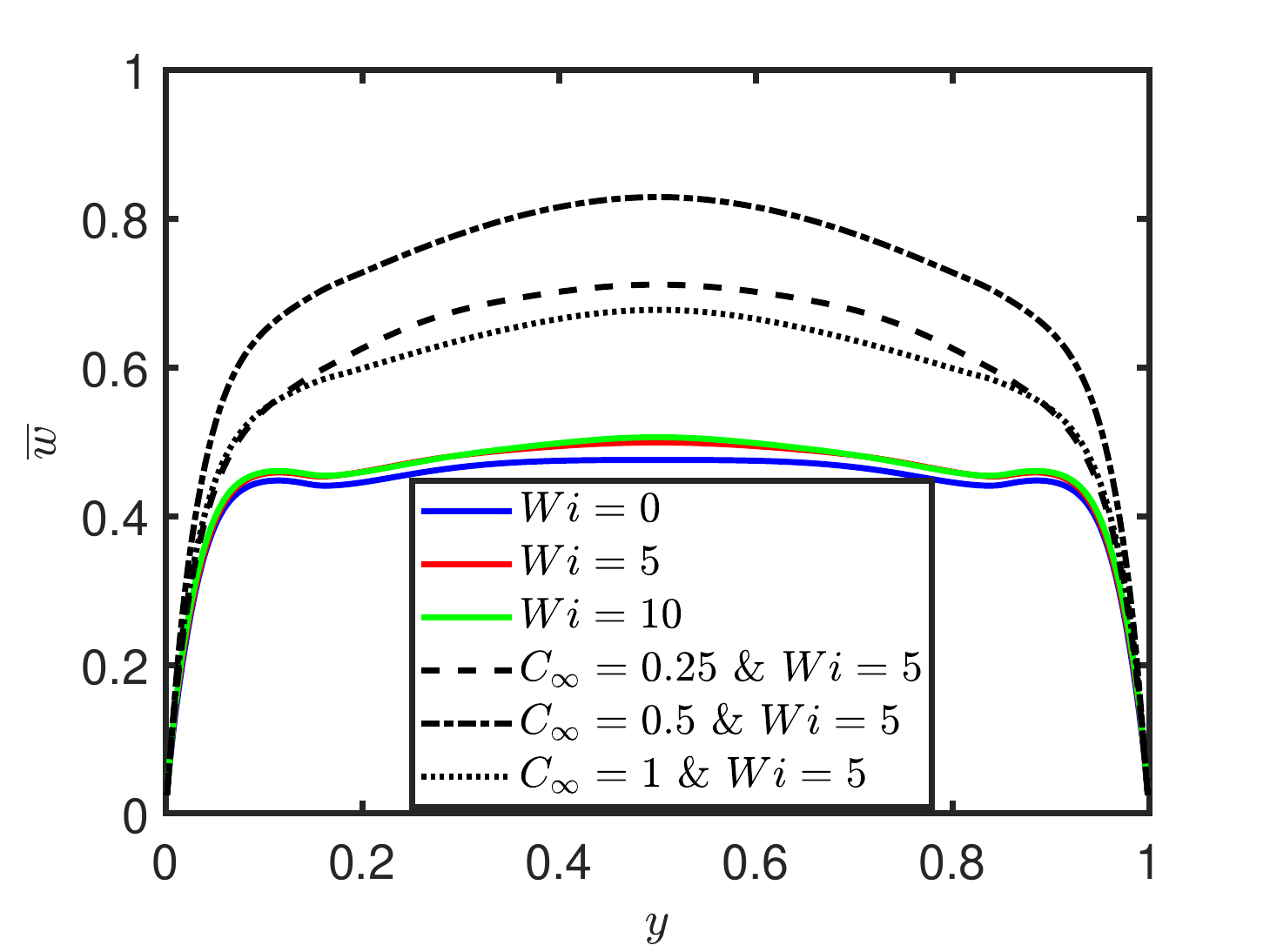}
\caption{}
\label{flow_rateV2}
\end{subfigure}
\caption{Newtonian flow (First row): (a) time evolution of the liquid flow rate for varying surfactant concentration, and (b) the averaged liquid velocity versus the wall-normal coordinate, after the flow reaches an approximate steady state. Non-Newtonian flow (second row):(c) time evolution of the liquid flow rate for a clean fluid and varying Weissenberg number, and for a fixed Weissenberg number and varying surfactant concentration (d) the averaged liquid velocity versus the wall-normal coordinate, after the flow reaches an approximate steady state. $(\overline{w}=\overline{w^*}/W^*_b)$}
\label{flow_rate}
\end{figure}

Simulations are first performed to examine the sole effects of soluble surfactant on the bubbly channel flow. For this purpose, the both phases are set to be Newtonian and the bulk surfactant concentration is varied in the range of $C_\infty = 0$ (clean) and $C_\infty =1$ (highly contaminated). Figure~\ref{flow_rateN}~\&~\ref{flow_rateN2}  presents the evolution of the liquid flow rate and the liquid velocity profiles at approximately statistically steady state for the clean and contaminated cases respectively. It can be seen that the presence of surfactant has drastic effect on the dynamics of the turbulent bubbly flow. The addition of nearly spherical bubbles to a statistically steady state turbulent single phase flow causes significant increase in the drag and thus the flow rate decreases in comparison to the corresponding single phase flow. The addition of surfactant in the bubbly flows prevents this reduction of the flow rate. The maximum prevention occurs for the $C_\infty=0.5$ case, for which the flow rate approaches to that of single phase flow rate in the statistically steady state. 

\begin{figure}[!htb]
\setlength{\unitlength}{1cm}
\begin{center}
\begin{tabular}[c]{ccc}
{\includegraphics[trim={2.5cm 0.1cm 3cm 0.1cm},clip,width=4.6cm]{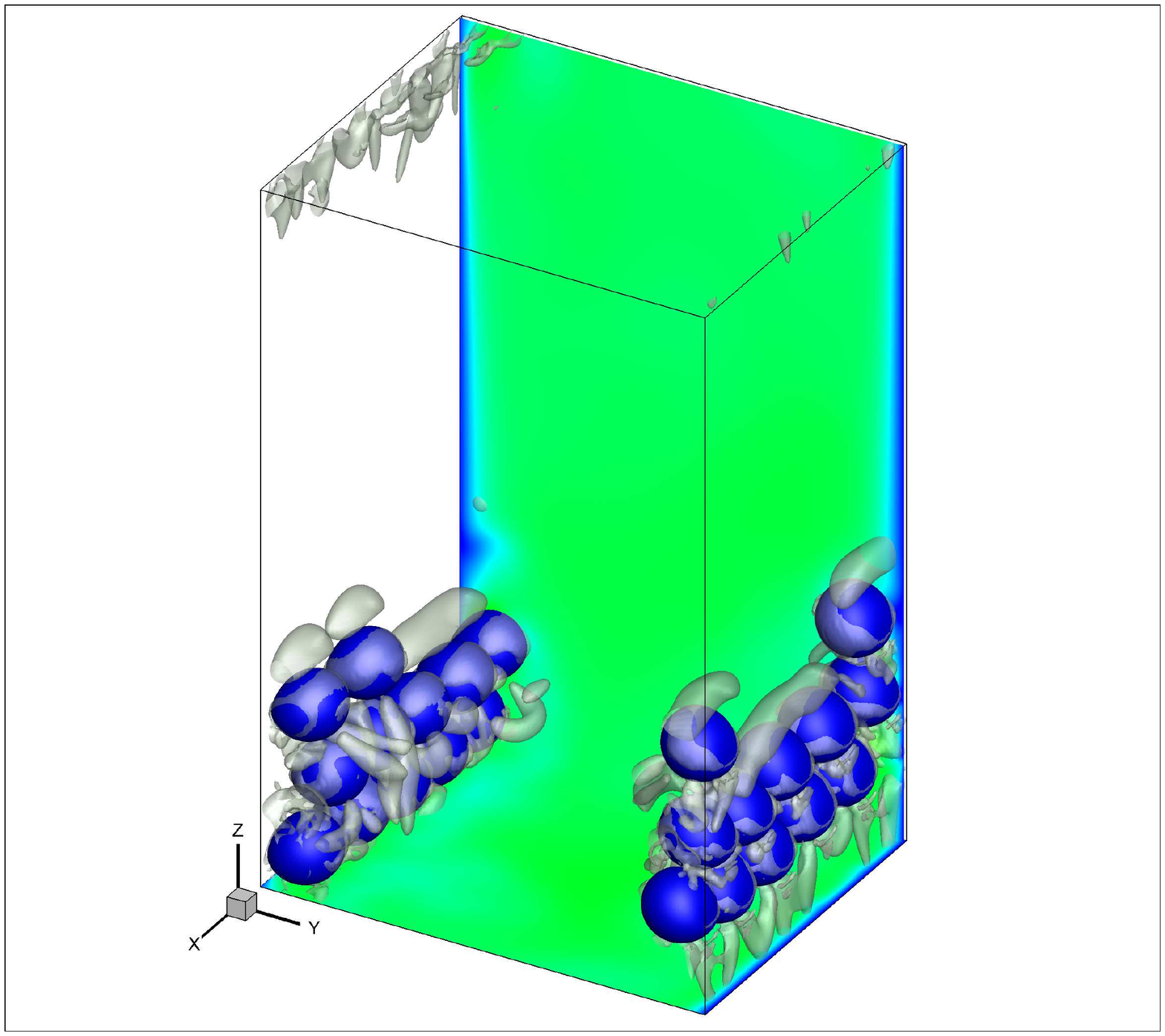}} &
\hspace{-1.0cm}
{\includegraphics[trim={3cm 1cm 3cm 0.1cm},clip, width=5cm]{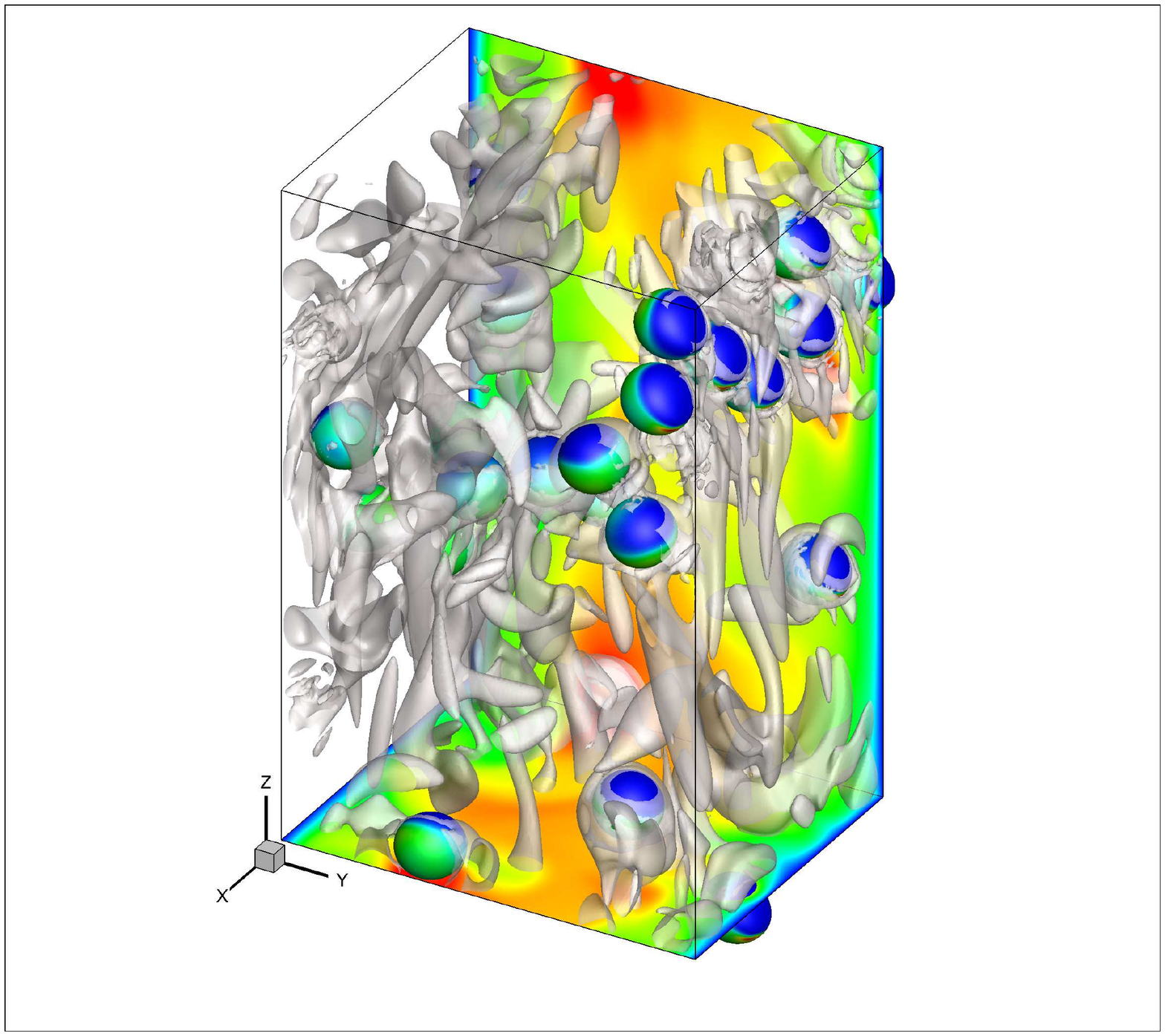}} &
\hspace{-1.3cm}
{\includegraphics[trim={3cm 1cm 3cm 0.1cm},clip, width=5cm]{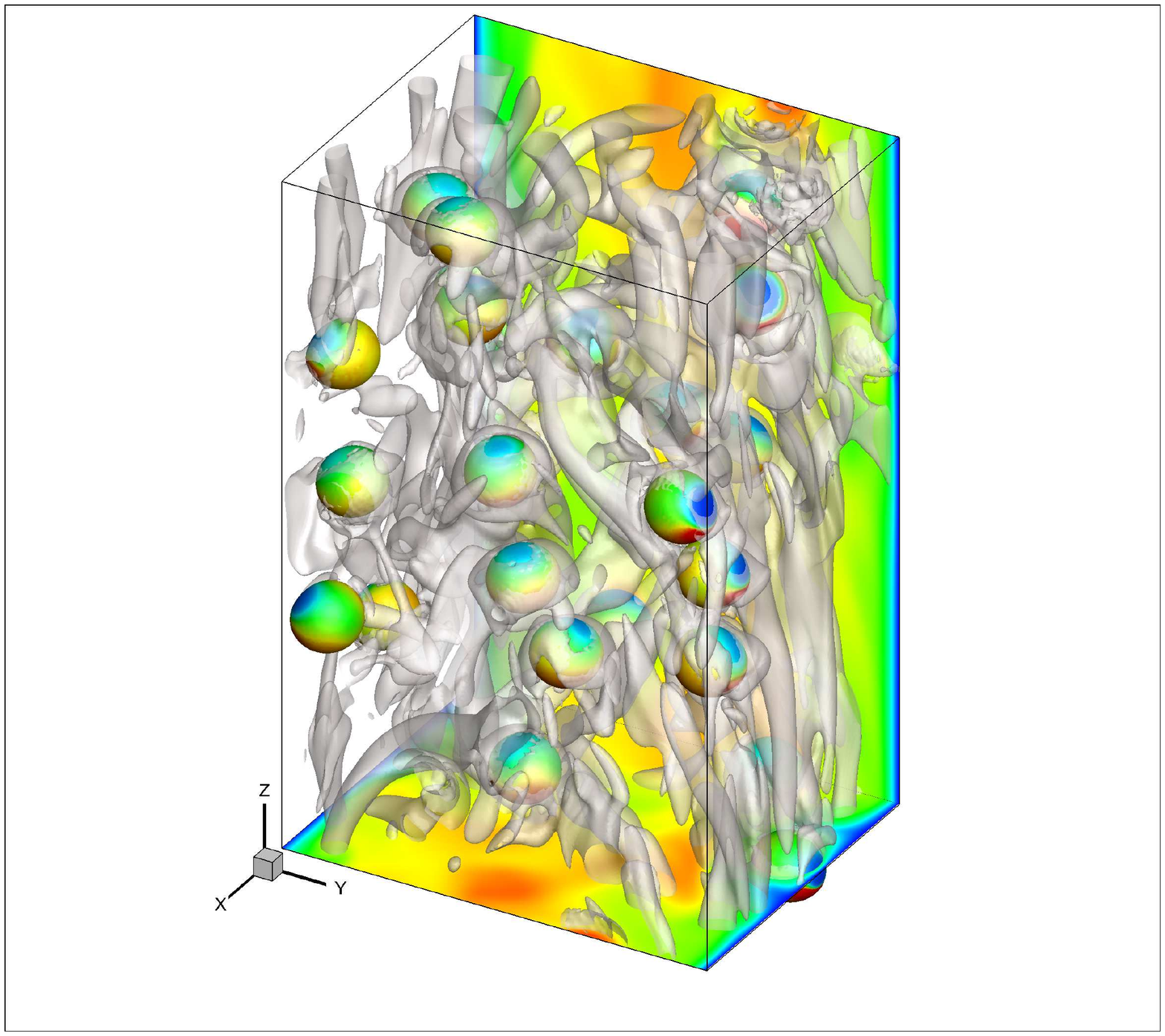}} \\  
(a) $C_{\infty}=0$ &  (b) $C_{\infty}=0.25$ & (c) $C_{\infty}=0.5$
\end{tabular}
\end{center}
\caption{Sole effects of surfactant on the vortical structures in the fully-developed regime. The vortical structures are visualized using $Q$-criterion, i.e., the iso-contours of the normalized second invariant of the velocity-gradient tensor, $Q^*/(W^*_b/(2h^*))^2=1$, are plotted. The contours on the plane represent instantaneous streamwise velocity with the scale ranging from $0.15W^*_b$ (blue) to $1.1W^*_b$ (red). The contours on the bubble surface represent interfacial surfactant concentration $(\Gamma)$ with the scale ranging from $0$ (blue) to $0.15$ (red).}
\label{figQ1}
\end{figure}

\begin{figure}[!htb]
\begin{tabular}[c]{c c c c}
  {\includegraphics[trim={0cm 0.1cm 0.3cm 0.0cm},clip,width=3.1cm]{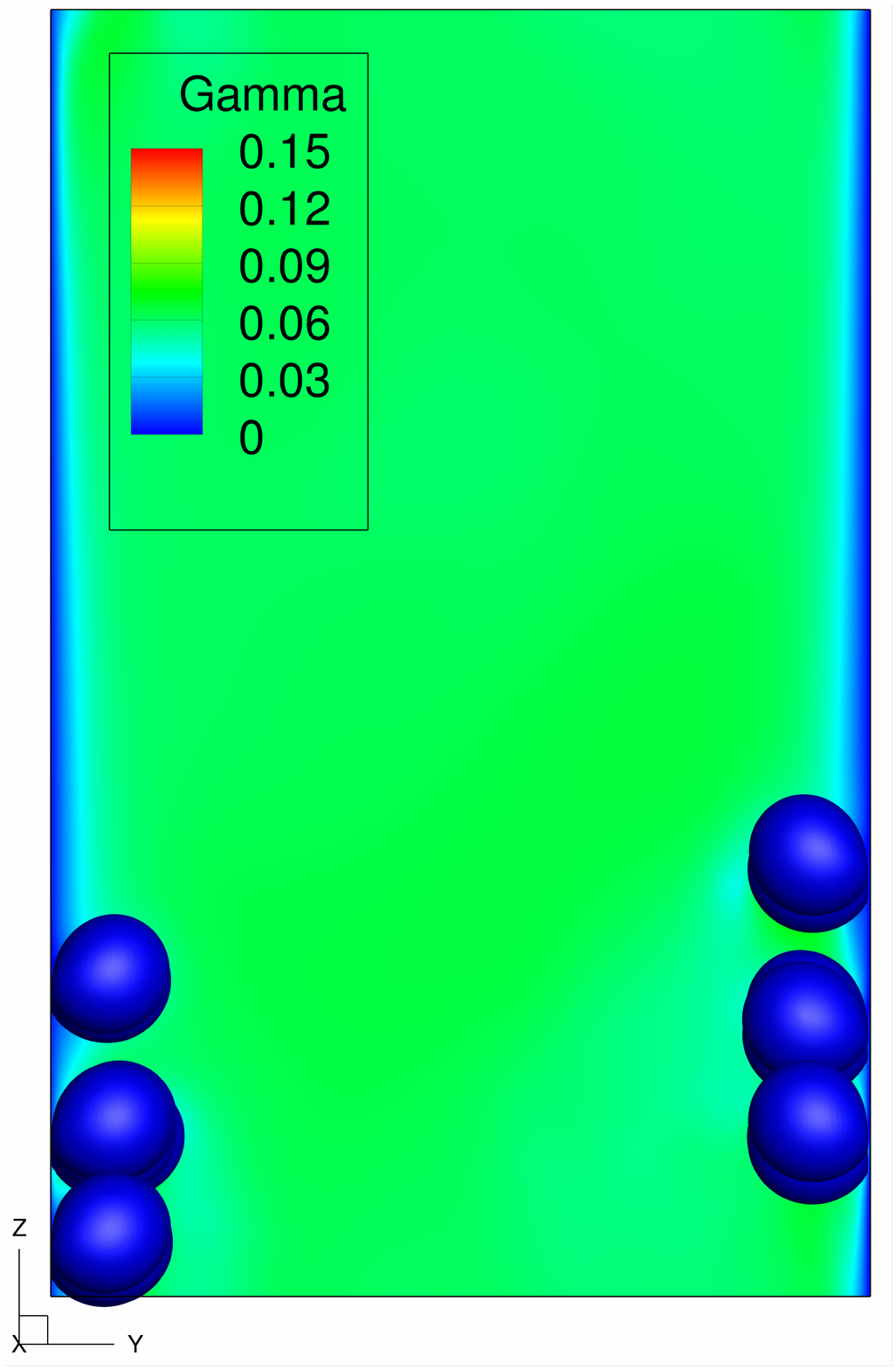}} &
  {\includegraphics[trim={0cm 0.1cm 0.3cm 0.0cm},clip,width=3.4cm]{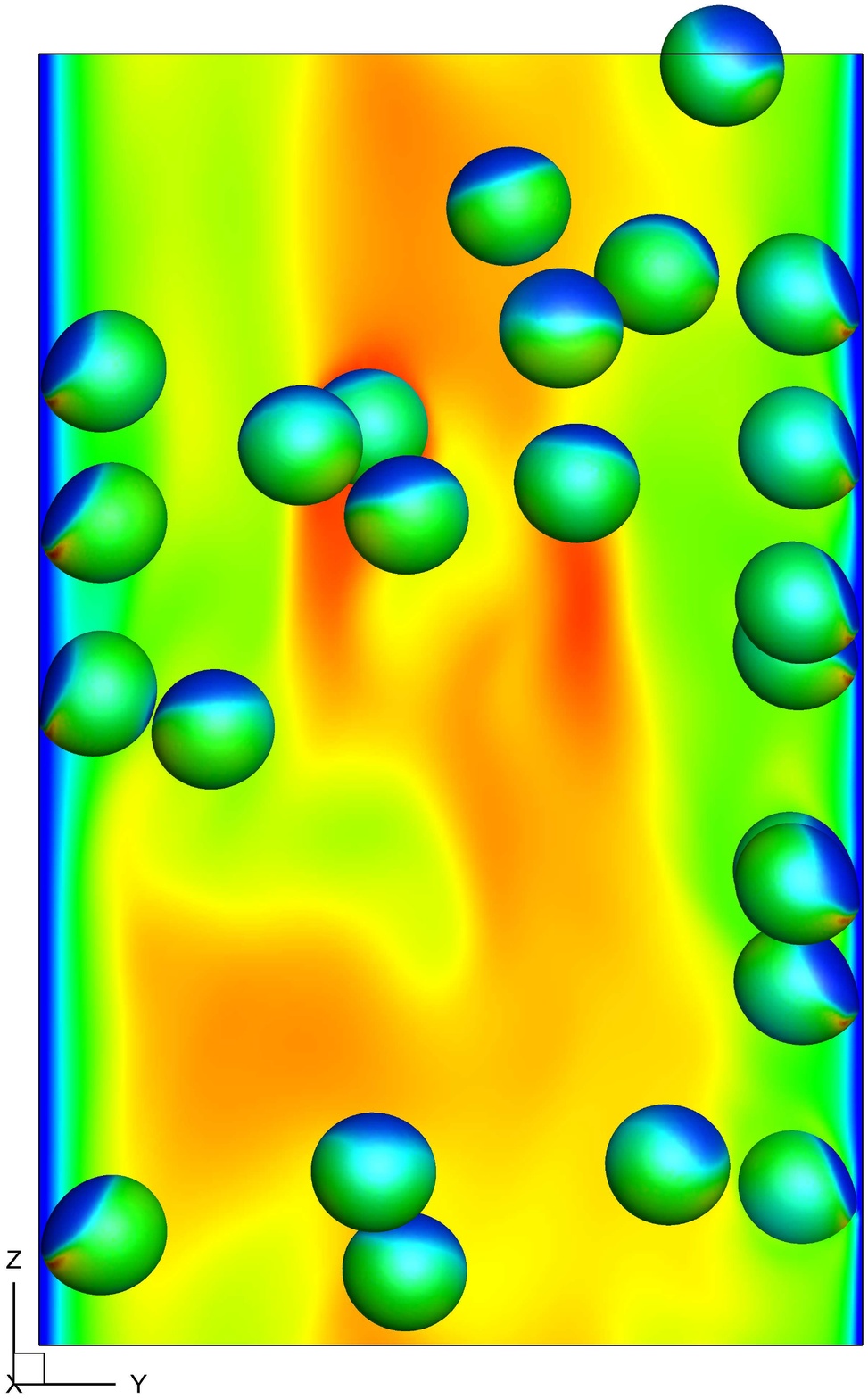}} &
  {\includegraphics[trim={0cm 0.1cm 0.3cm 0.0cm},clip,width=3.4cm]{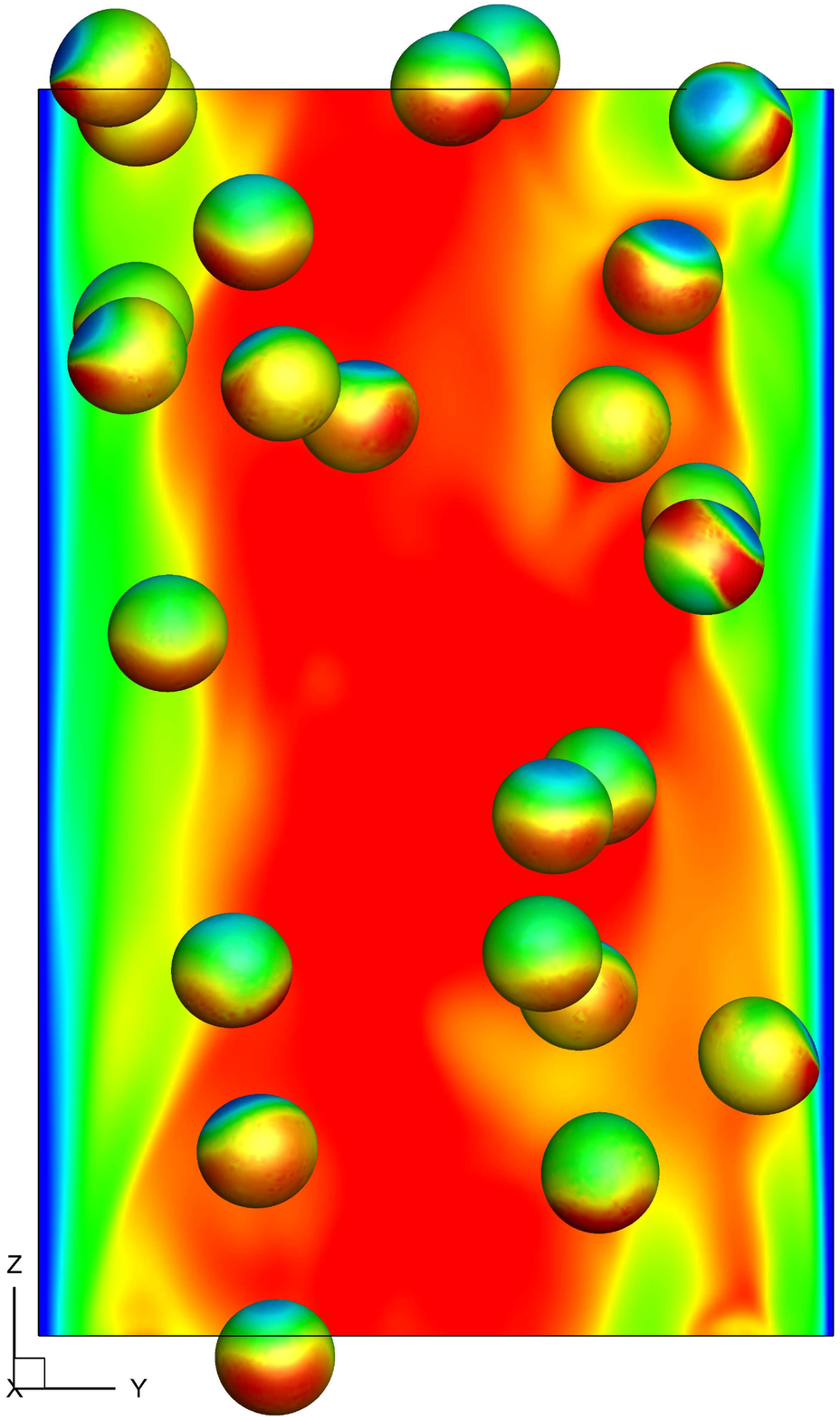}} &
  {\includegraphics[trim={0cm 0.1cm 0.3cm 0.0cm},clip,width=3.4cm]{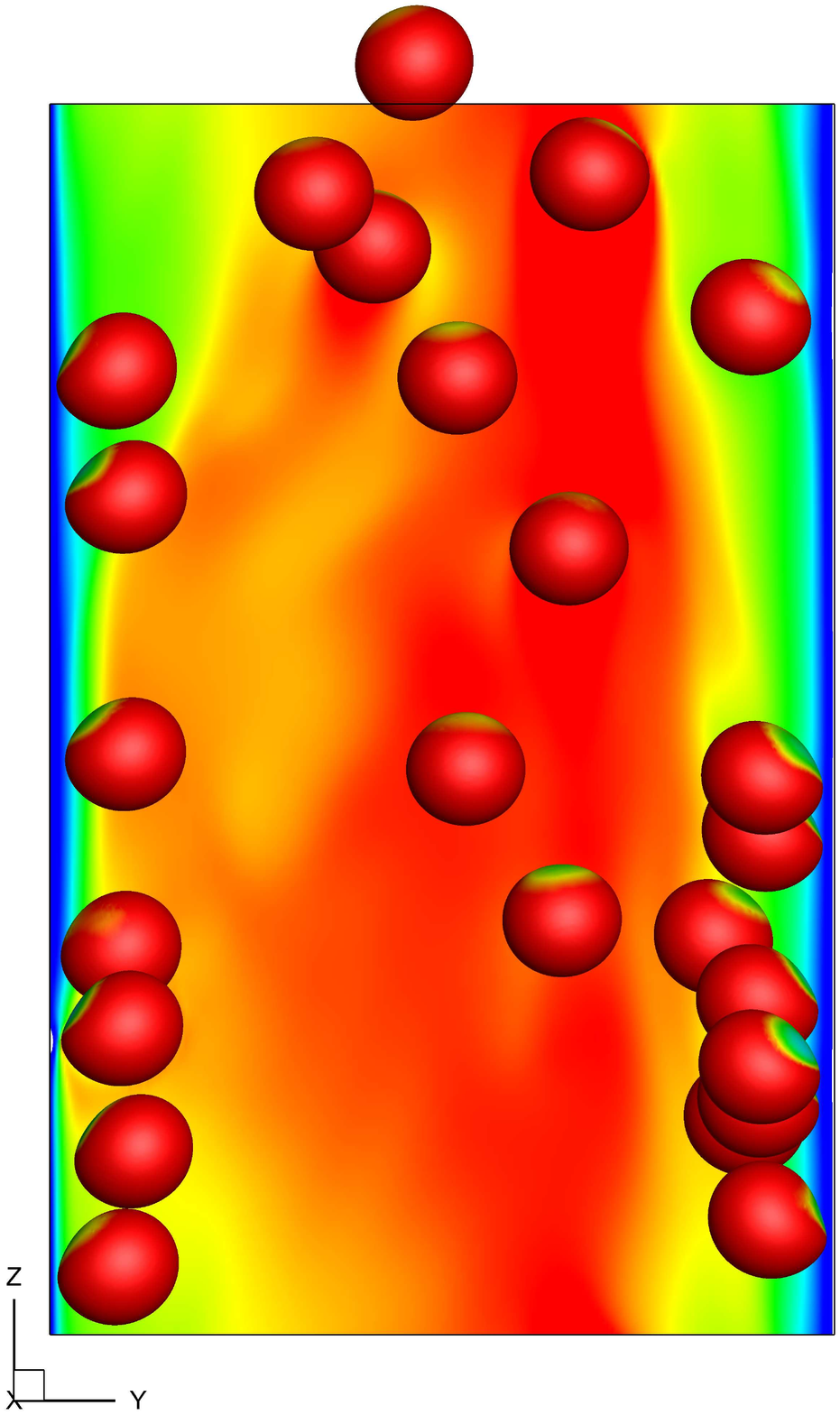}}\\
{\includegraphics[trim={0cm 0.1cm 0.3cm 0.0cm},clip,width=3.1cm]{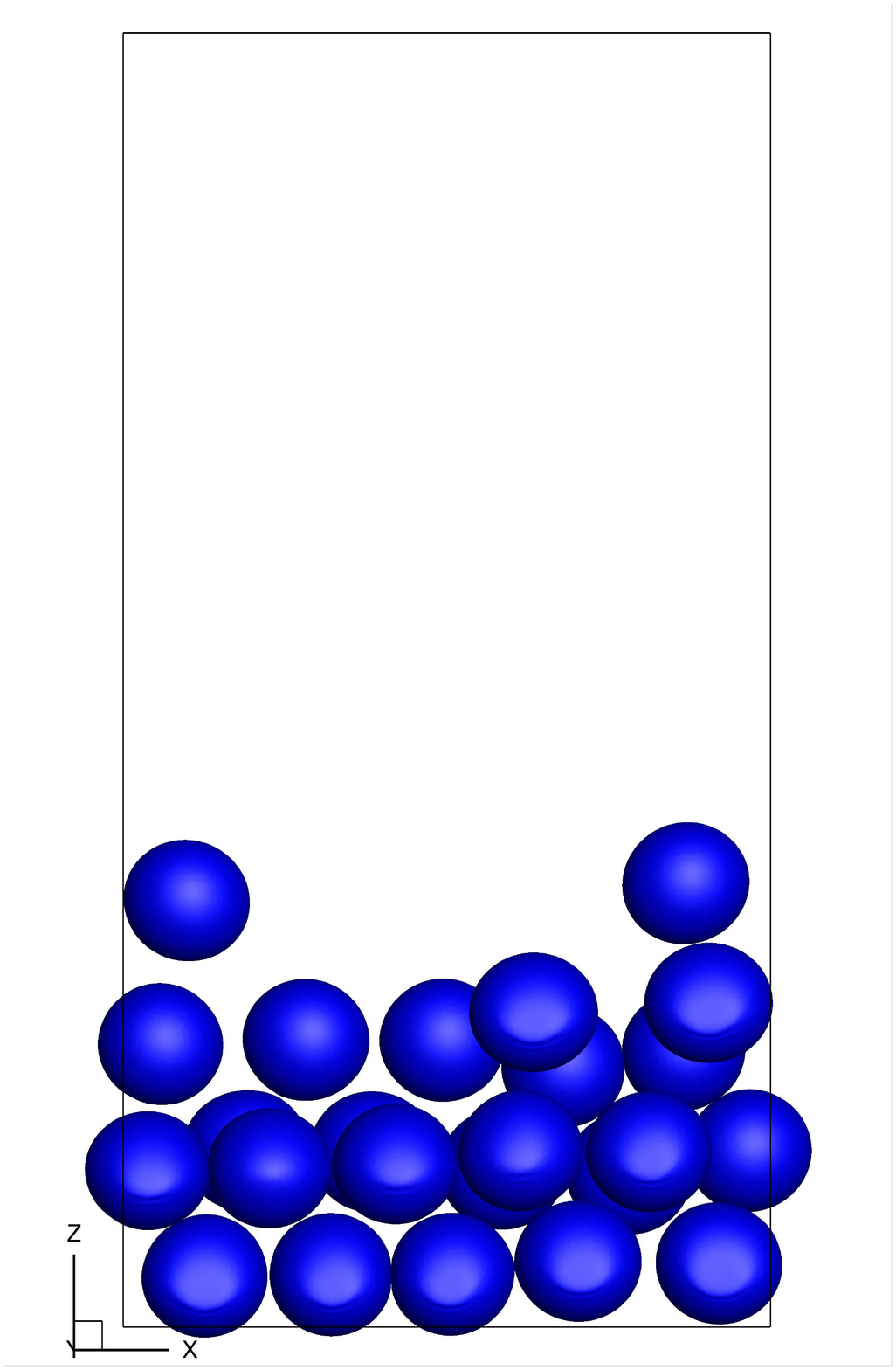}} &
{\includegraphics[trim={0cm 0.1cm 0.3cm 0.0cm},clip,width=3.4cm]{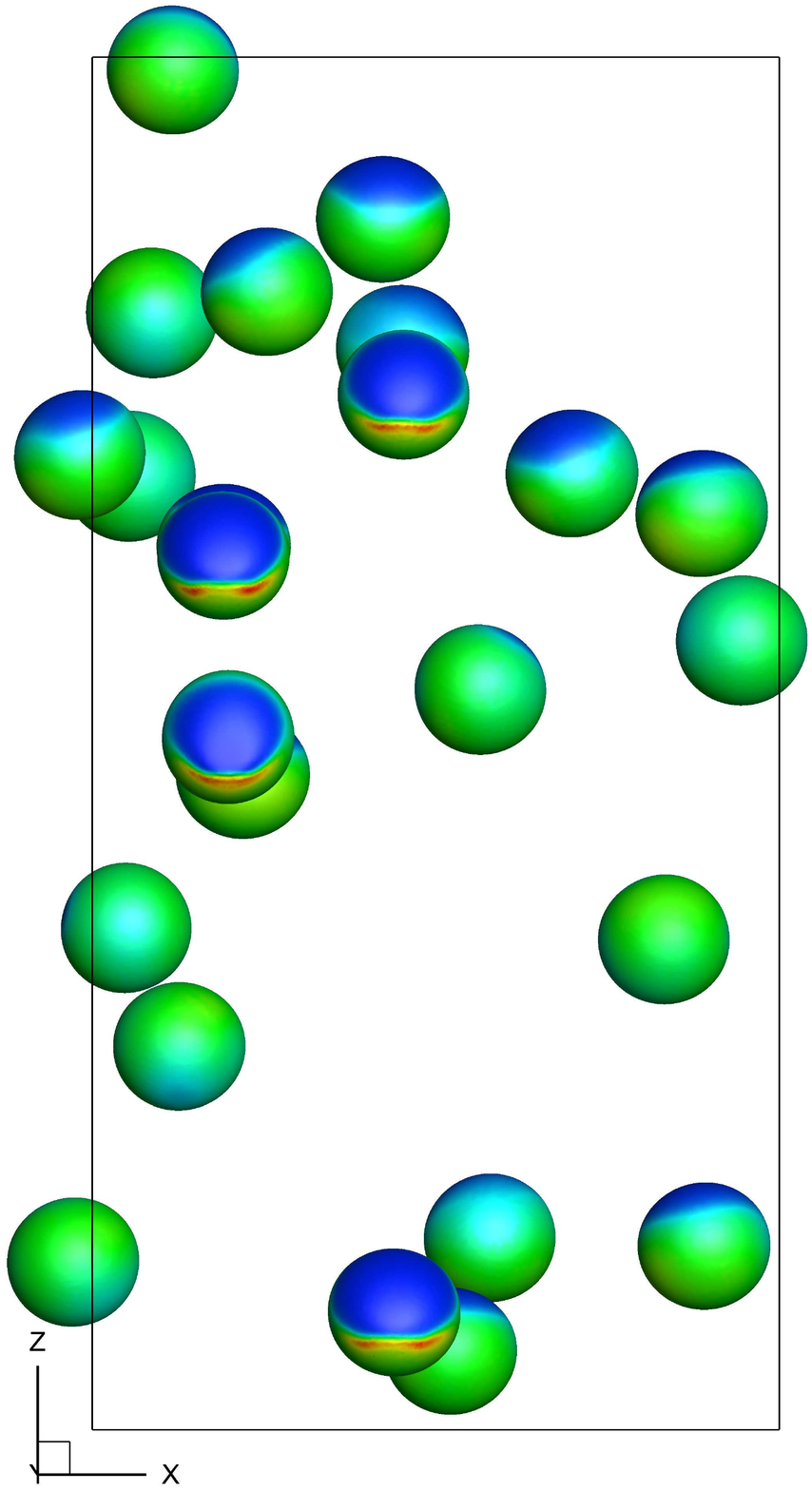}} &
{\includegraphics[trim={0cm 0.1cm 0.3cm 0.0cm},clip,width=3.4cm]{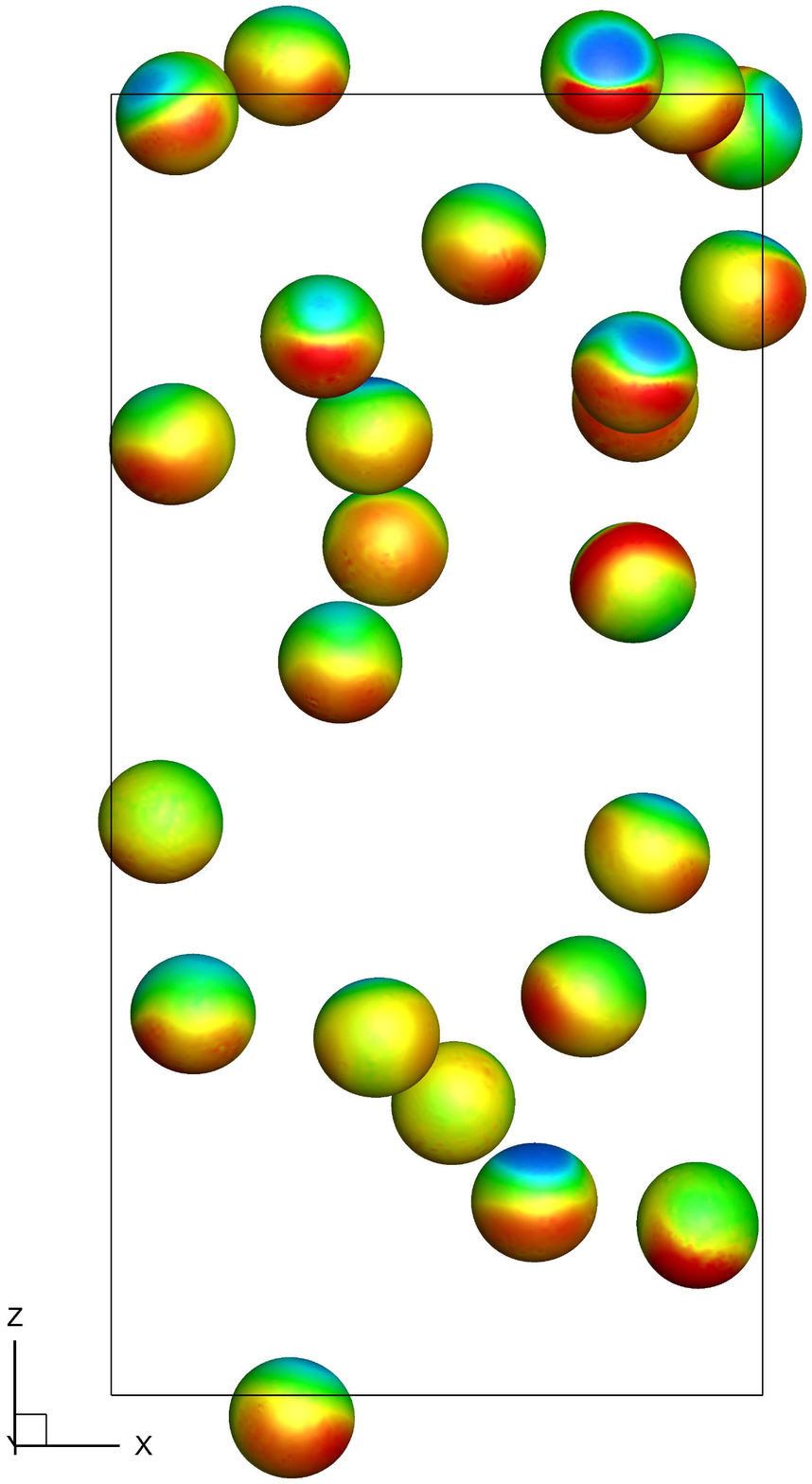}} &
{\includegraphics[trim={0cm 0.1cm 0.3cm 0.0cm},clip,width=3.4cm]{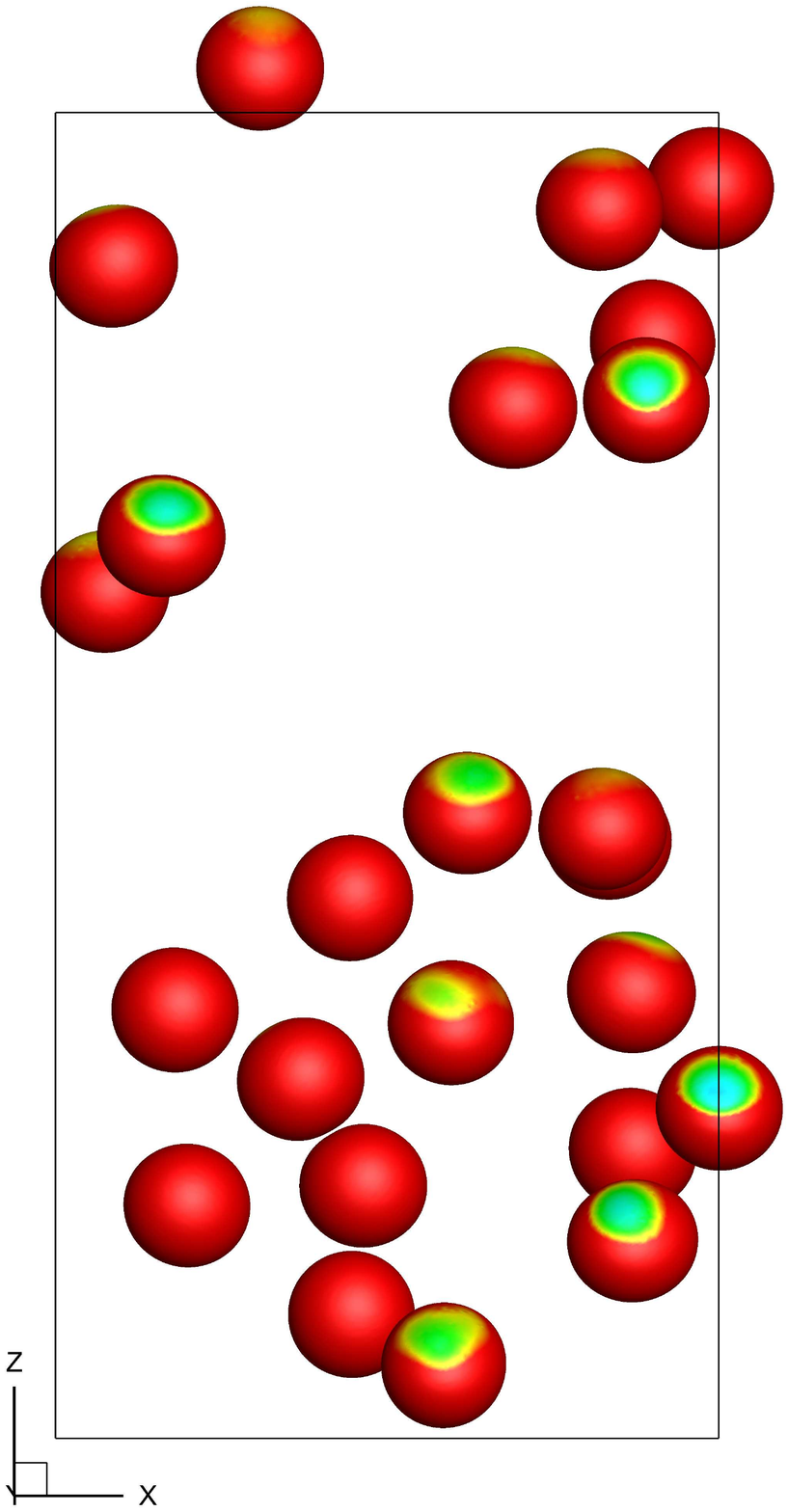}}\\                  
     $C_{\infty}=0$  &   $C_{\infty}=0.25$  & $C_{\infty}=0.5$ & $C_{\infty}=1.0$
\end{tabular}
\caption{Sole effects of surfactant on the structure of Newtonian turbulent bubbly flow: bubble distribution on the streamwise-wall-normal plane (first row) and on the streamwise-spanwise plane (second row) when the flow has reached a statistically steady state. The contours on the plane represent instantaneous streamwise velocity with the scale ranging from $0.15W^*_b$ (blue) to $1.1W^*_b$ (red).}
	\label{BubbledistN}
\end{figure}

Figure~\ref{figQ1} shows the vortical structure of the bubbly flow for the bulk surfactant concentrations of $C_\infty= 0$, $0.25$ and $0.5$. The vortical structures are identified using the $Q$-criterion~\cite{jeong1995identification} (see also the figure caption). Clearly, the presence of contamination exhibits a drastic effect on the flow. For a clean suspending phase, the dispersed bubbles migrate toward the wall and seem to relaminarize the flow -- the only regions of high vorticity are associated with the footprint of the bubble wakes. With an increasing surfactant concentration, the bubbles appear to be more homogeneously distributed, and the flow is clearly in a turbulent state. The constant contours of the interfacial surfactant concentration and distribution of bubbles in the streamwise-wall-normal and streamwise-spanwise plane  are shown in Fig.~\ref{BubbledistN}. For the clean case, the bubbles migrate toward the channel wall mainly due to the hydrodynamic lift force~\citep{Lu06}. When surfactant is  added, the Marangoni stresses counteract this lift force and eventually reverse the direction of the lateral migration of the bubbles when $C_\infty$ exceeds a critical value. As $C_\infty$ is increased from $0$ to $0.25$, the number of bubbles in the core region is increased, though we still can see the wall-layer. In this case, the amount of surfactant adsorbed on the surface of the bubbles is significant, but not large enough to completely overcome the hydrodynamic lift that is responsible for the bubble wall layering. With a further increase of $C_\infty$, a larger portion of the area of the bubbles is contaminated. As a result, all bubbles detach from the wall, migrate toward the core region and spread out the channel cross-section, drastically changing the dynamics and the entire structure of the bubbly flows, as seen in Figs.~\ref{flow_rateN} and \ref{BubbledistN}. 

\begin{figure}[!htb]
\setlength{\unitlength}{1cm}
\begin{center}
\begin{tabular}[c]{cc}
{\includegraphics[width=7cm]{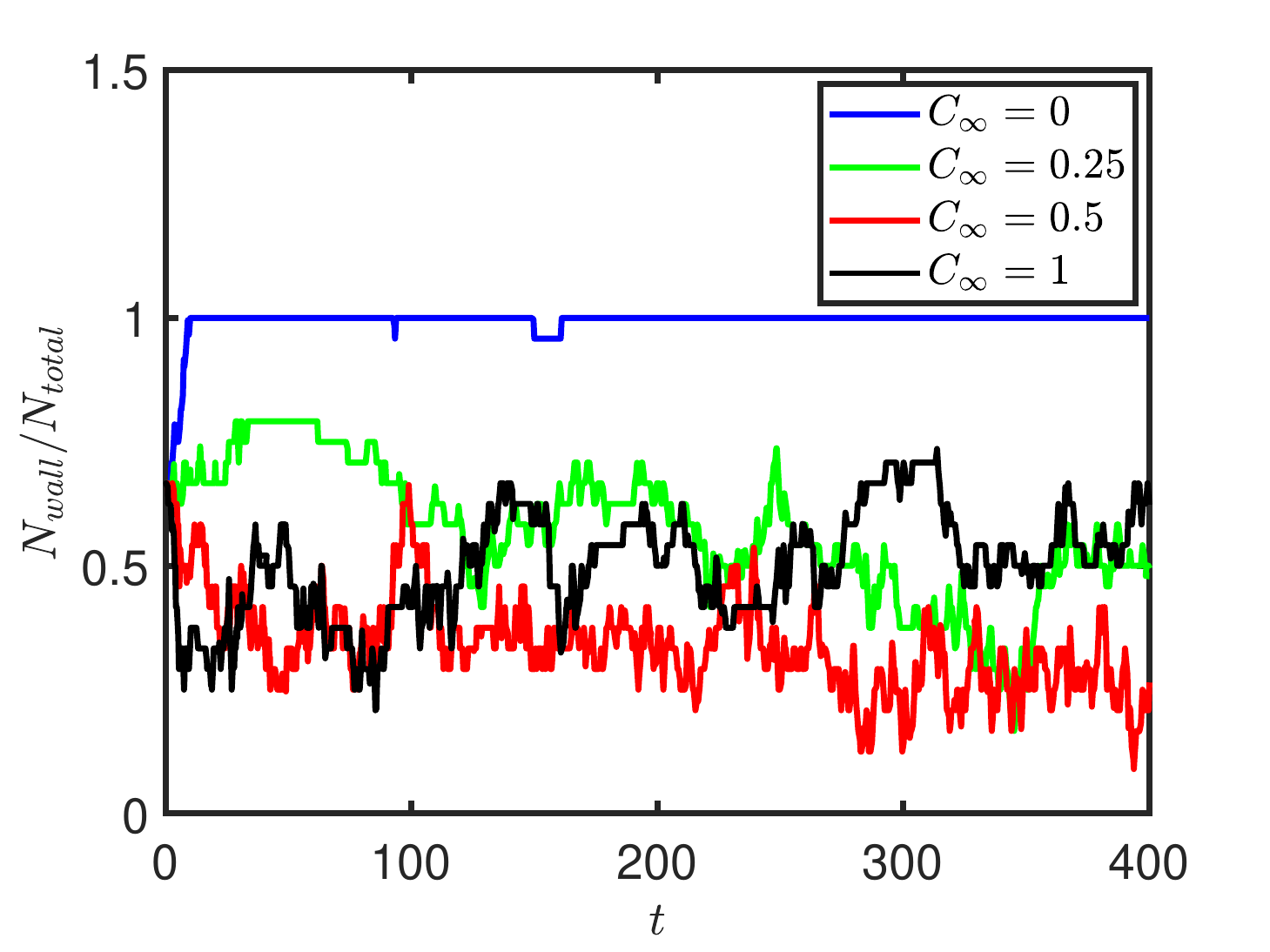}}&
{\includegraphics[width=7cm]{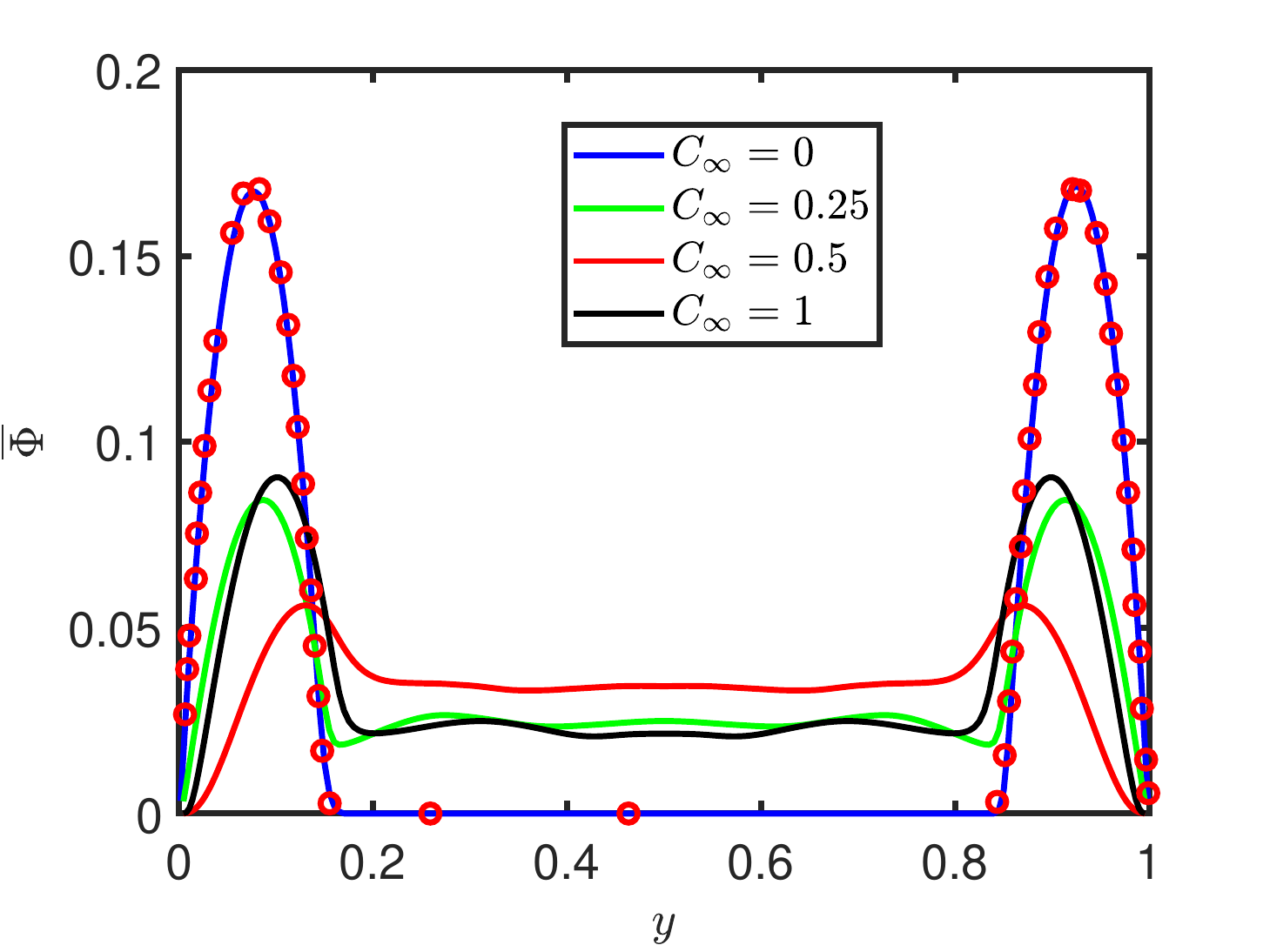}} \\
{\includegraphics[width=7cm]{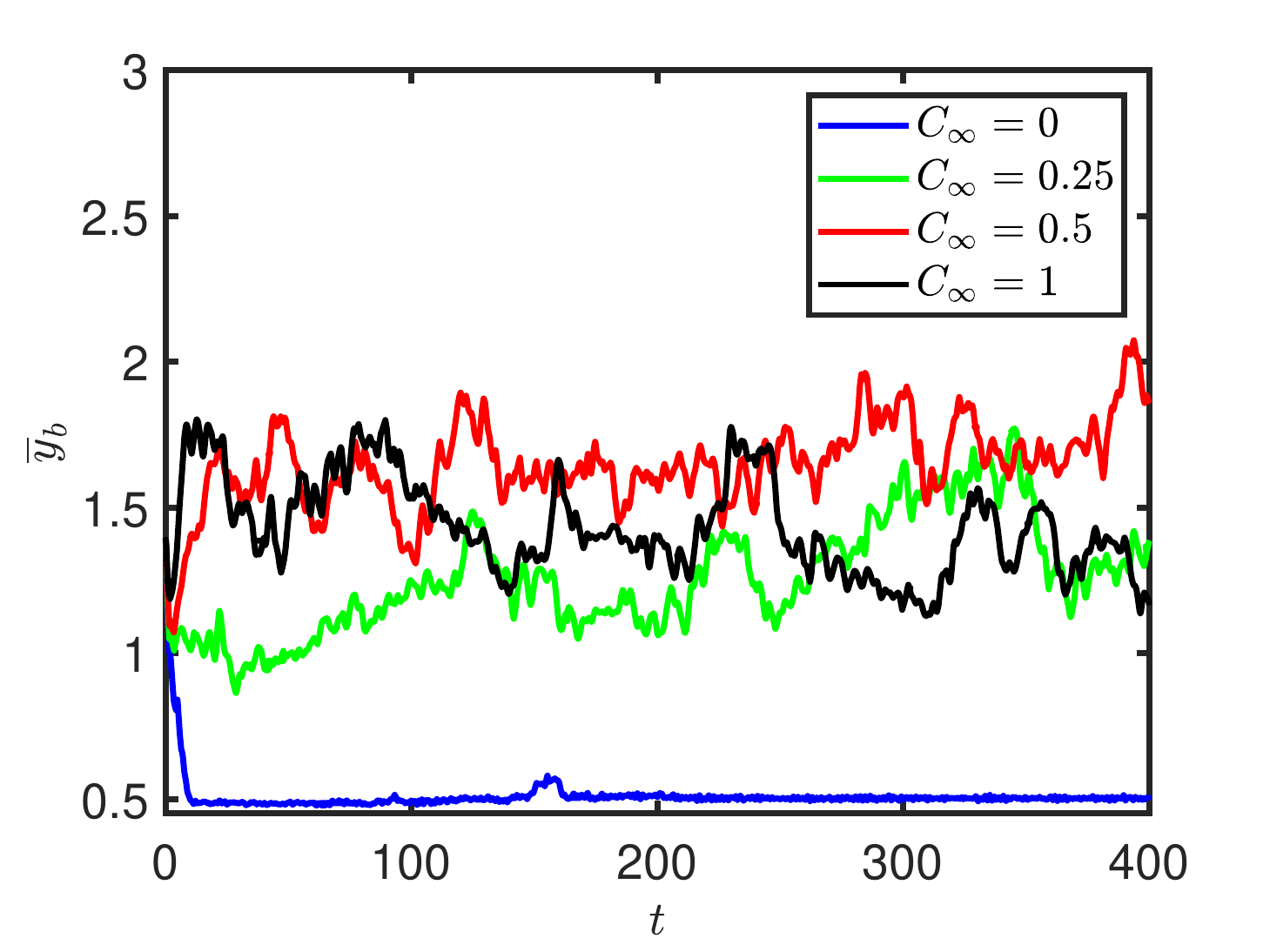}}&
{\includegraphics[width=7cm]{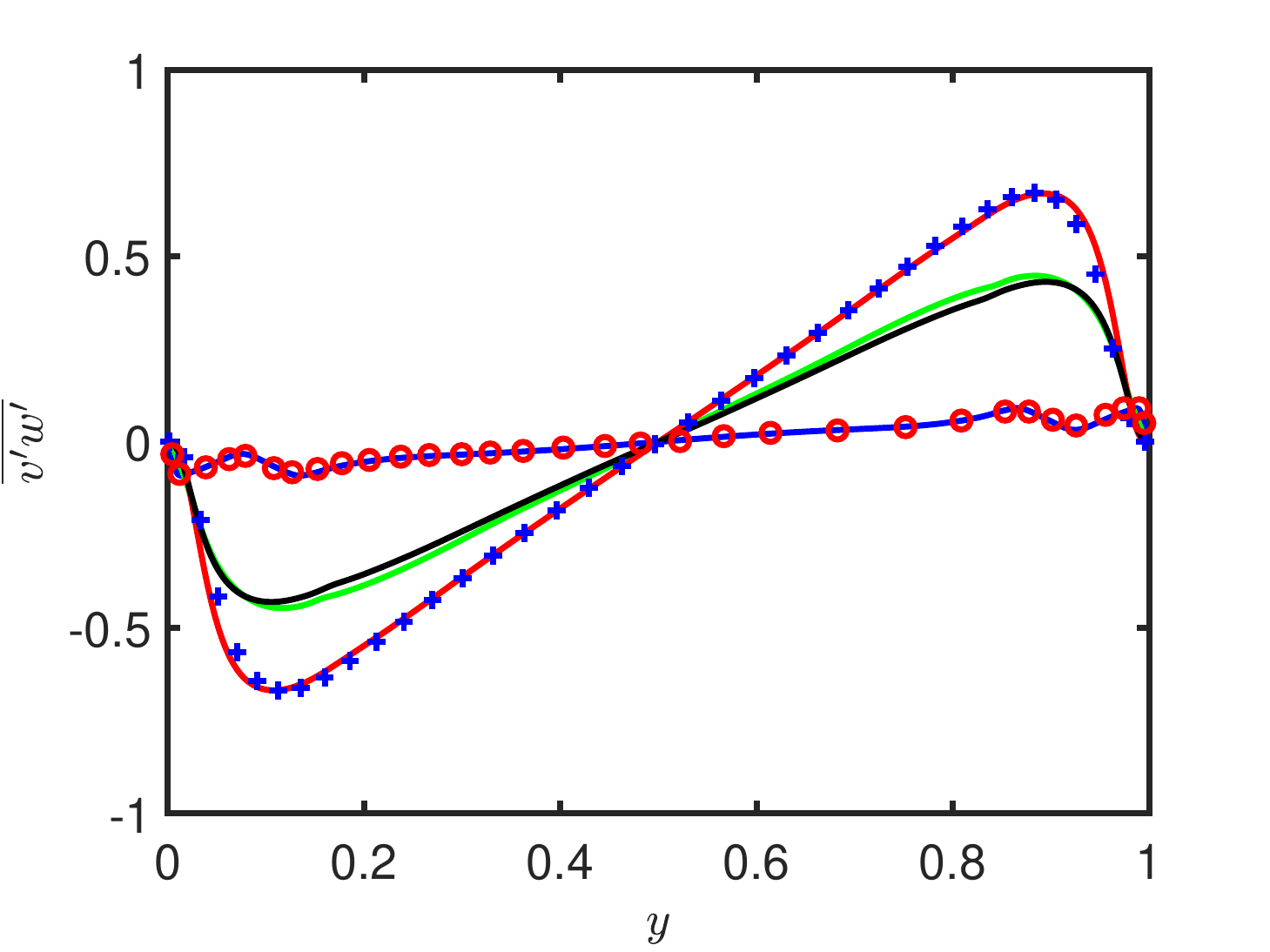}} \\
{\includegraphics[width=7cm]{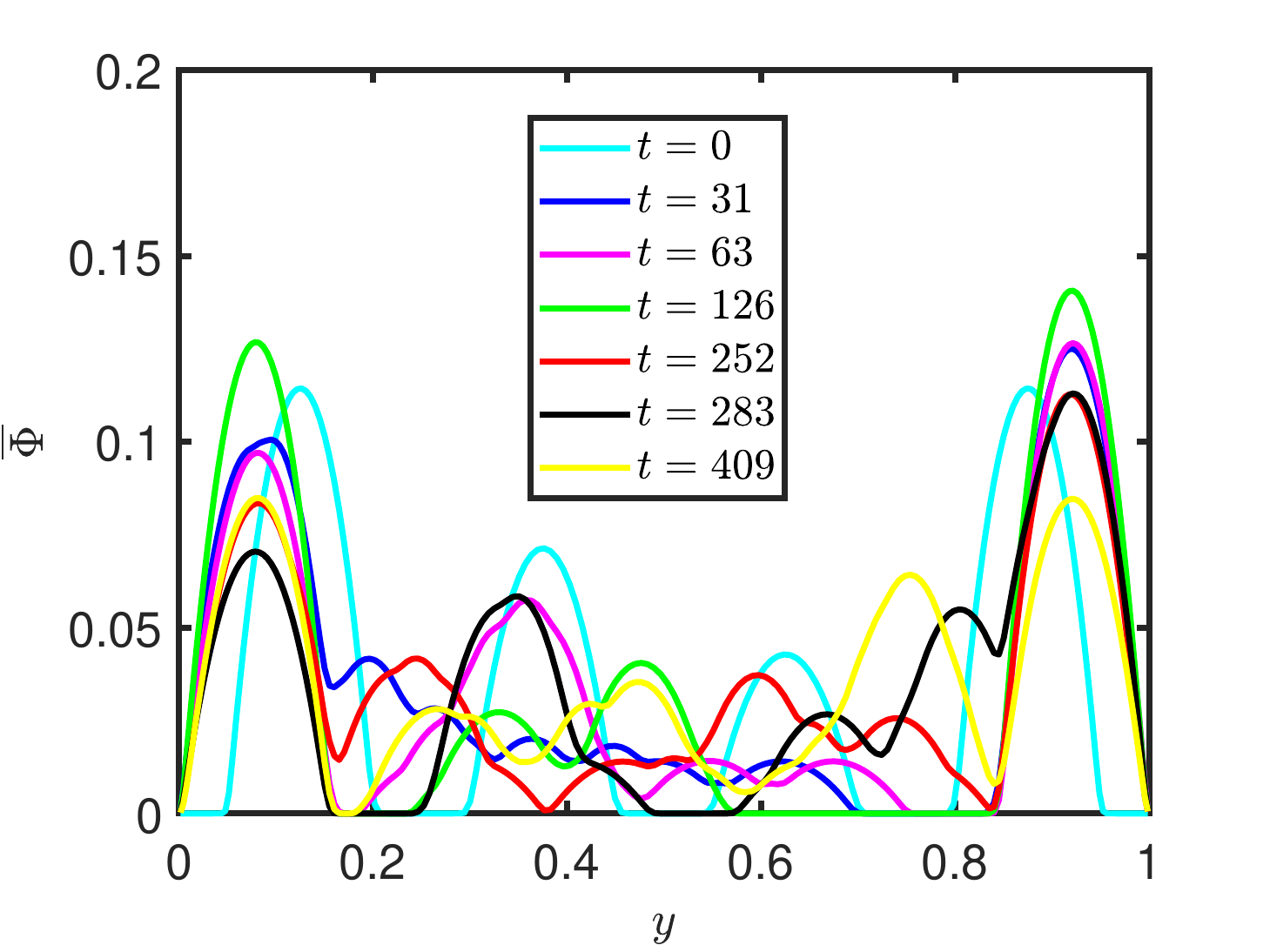}}&
{\includegraphics[width=7cm]{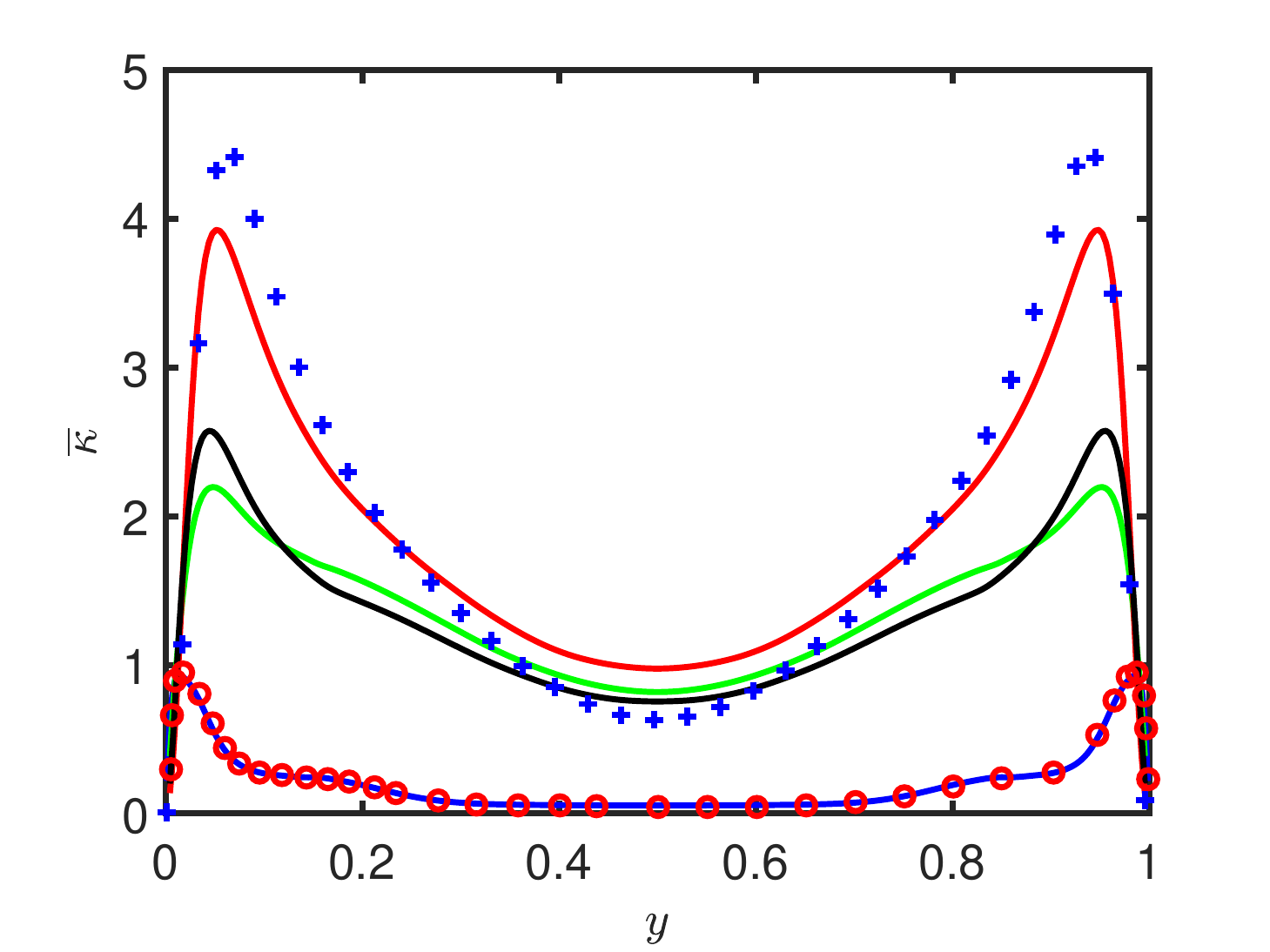}} \\
\end{tabular}
\end{center}
\vspace{-1cm}
\caption{Sole effects of surfactant. The first and second columns show the transient and statistically stationary results, respectively. In the left column, the evolution of the number of bubbles in the wall layer (top), the average distance of the bubbles from the wall $(\overline{y_b}=\overline{y^*_b}/d^*_b)$ (middle) and the average void fraction at various times for $C_\infty=0.25$  (bottom). In the right column, statistically steady state results of the void fraction (top), the Reynolds stresses scaled by ${w^*_\tau}^2$ (middle) and the turbulent kinetic energy scaled by ${w^*_\tau}^2$ (bottom).}
\label{statssrf}
\end{figure}

To quantify the effects of surfactants on the flow, the transient and time-averaged quantities are shown in Fig.~\ref{statssrf}. The time-averaged quantities are obtained by averaging over a period of $200$ computational time units once the flow reaches a statistically steady state. Note that the time unit is defined as $h^*/W^*_b$. The results show that the bubbly wall layer formation in the clean case has thickness of a bubble diameter as seen in the evolution of the number of bubbles in the wall layer (i.e., the number of bubbles with a centroid position lower than one bubble diameter). It is interesting to note that, for the clean bubbly flows, the formation of bubble-rich wall-layer with a thickness of about a bubble diameter is also observed in the laminar~\citep{Lu06}, weakly turbulent~\citep{Lu} and also for highly turbulent flows~\citep{Lu13}. For the contaminated cases, as $C_\infty$ increases, the bubbles move toward the core region. The void fraction peaks are less pronounced and located further away from the walls. For the highest concentration of $C_\infty=1$, wall-layers again emerge similar to the case with $C_\infty=0.25$. This interesting behavior is attributed to the fact that the bubble surface is nearly uniformly covered by the surfactant at the equilibrium concentration, resulting in a nearly uniform reduction in surface tension without much Marangoni stresses in the case of excessive surfactant concentrations. As regards the Reynolds stresses, they increase from almost zero in the bulk of the channel for the clean case to the almost linear profile resembling the characteristic of a single-phase turbulent flow for the contaminated cases. As $C_\infty$ increases, the Reynolds stress profile becomes closer  and eventually matches that of the single-phase turbulent flow. It can be seen that, for the clean case, the turbulent kinetic energy level is low, which is consistent with the observation that the flow relaminarizes for the clean case. But the values of the kinetic energy are increased with $C_\infty$, except for the case of $C_\infty=1$. \citet{Lu} proposed that increase in the kinetic energy could be due to enhanced vorticity generation created by surfactant-induced rigidification of interfaces. The decrease in the Reynolds stress and turbulent kinetic energy are related to the increase in the bubble clustering near the wall. 

\subsection{Effects of viscoelasticity}\label{Veffects}

We next examine the effects of viscoelasticity on the turbulent bubbly channel flows. For this purpose, simulations are first performed for the Weissenberg numbers of $Wi = 0$ (Newtonian), $Wi = 5$ (moderately viscoelastic) and $Wi = 10$ (highly viscoelastic) in the absence of surfactant to demonstrate the sole effects of the visscoelasticity. Simulations are then performed to examine the combined effects of surfactant and viscoelasticity for $Wi = 5$ in the presence of surfactant with the bulk concentrations of  $C_{\infty}$ = 0.25, 0.5 and  1.0. The results are shown in  Fig.~\ref{flow_rateV}~\&~\ref{flow_rateV2}. As can be seen, for the clean viscoelastic case, the effect of polymers on the structure of the bubbly channel flow is opposite to that of surfactants. Increasing the Weissenberg number ($Wi$) promotes the formation of bubbly wall-layers and consequently reduction in the flow rate as shown in Fig.~\ref{BubbledistV} and Fig.~\ref{flow_rateV}, respectively. This interesting result is in fact consistent with the detailed numerical simulations of \citet{mukherjee2013effects} who showed that the matrix viscoelasticity induces a net force on the droplet pushing it towards the channel wall. As can be seen, at the initial stage ($t<30 \approx 2\lambda_{Wi=5}$), the viscoelastic cases follow the Newtonian case. This can be attributed to the fact that polymers react to the flow at a finite timescale ($\sim \lambda$). Initially, the polymer stresses are low, since there is not enough time for them to stretch. Eventually, the polymers fully stretch and the polymer stresses become significant. At about $t = 60 \approx 2\lambda_{Wi=10}$, the time evolution of the flow rate starts to diverge from the Newtonian one, but still follows its trend, i.e., decreases monotonically. For time $t >60$, two viscoelastic curves also separate due to different polymer stretching. It is well known that an addition of polymers results in drag reduction in single-phase flow~\cite{White}, however its effects are more complex in multiphase flows. The drag is determined by the interplay between the viscoelastic effects (drag reduction) and the formation of bubbly wall-layers (drag promotion).
In viscoelastic multiphase flows, the elastic lift is induced due to non-uniform normal stress difference \citep{mukherjee2013effects}. This elastic lift force move the bubbles toward the channel wall. Figure~\ref{BubbledistV} shows the bubble distribution and normalized first normal stress difference $(N_1)$ for $Wi=5$. As can be seen, the normal stress difference is larger on the wall side supporting the idea that the bubbles are forced to move towards the wall by the viscoelastic stresses. It is also seen that, for the clean case, the bubbles are accumulated at the walls forming horizontal bubble clusters. Bubble wall-layers are formed for all the Weissenberg numbers considered in this study. The formation of these wall-layers significantly diminishes the drag reduction effect of viscoelasticity observed for single-phase flows. 

\subsection{Combined effects of viscoelasticity and surfactant}\label{SrfV}
\begin{figure}[!htb]
\begin{center}
\begin{tabular}{c c c }
  {\includegraphics[trim={0cm 0.1cm 0.3cm 0.0cm},clip,width=4.4cm]{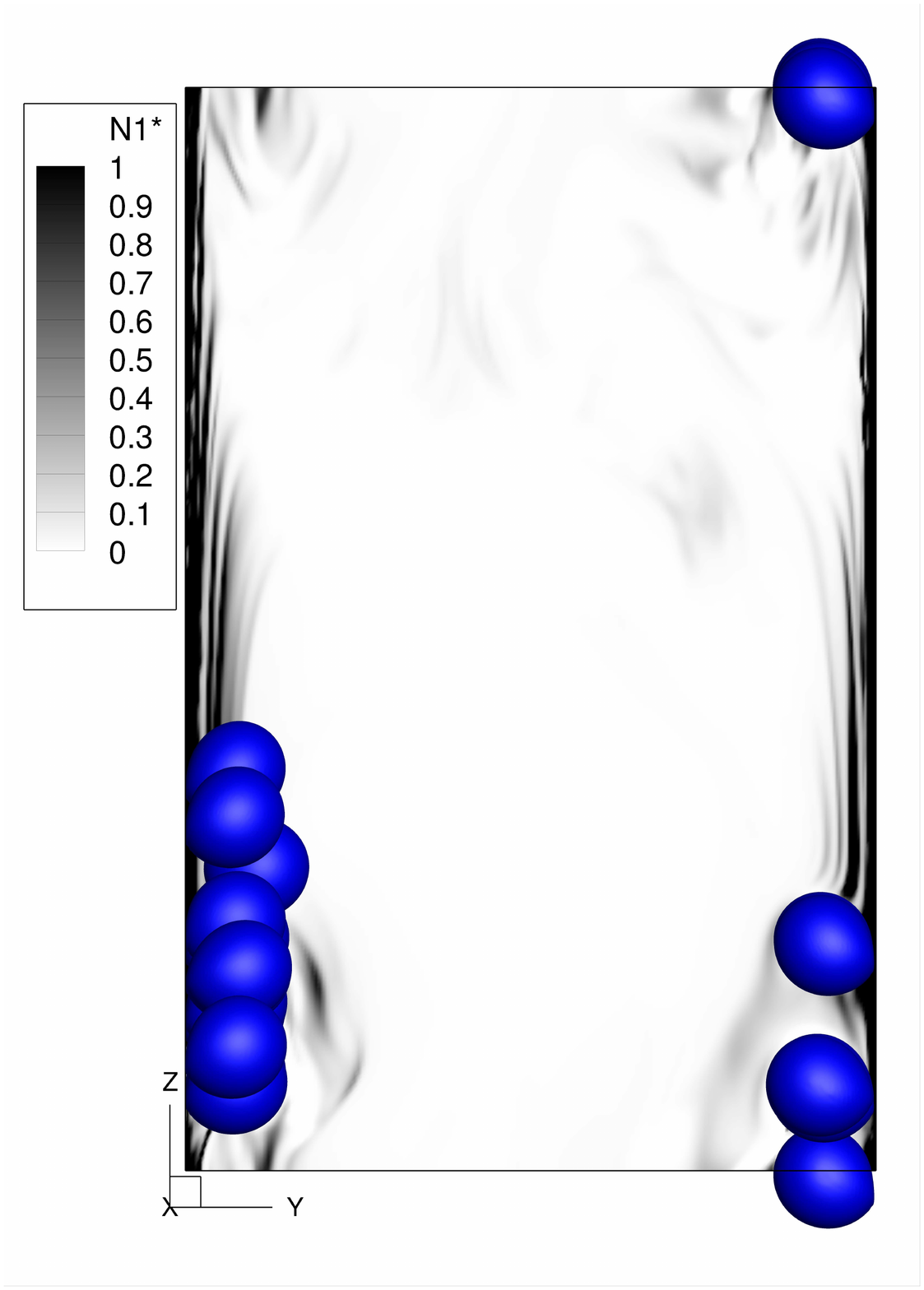}}  &
  {\includegraphics[trim={0cm 0.1cm 0.3cm 0.0cm},clip,width=4cm]{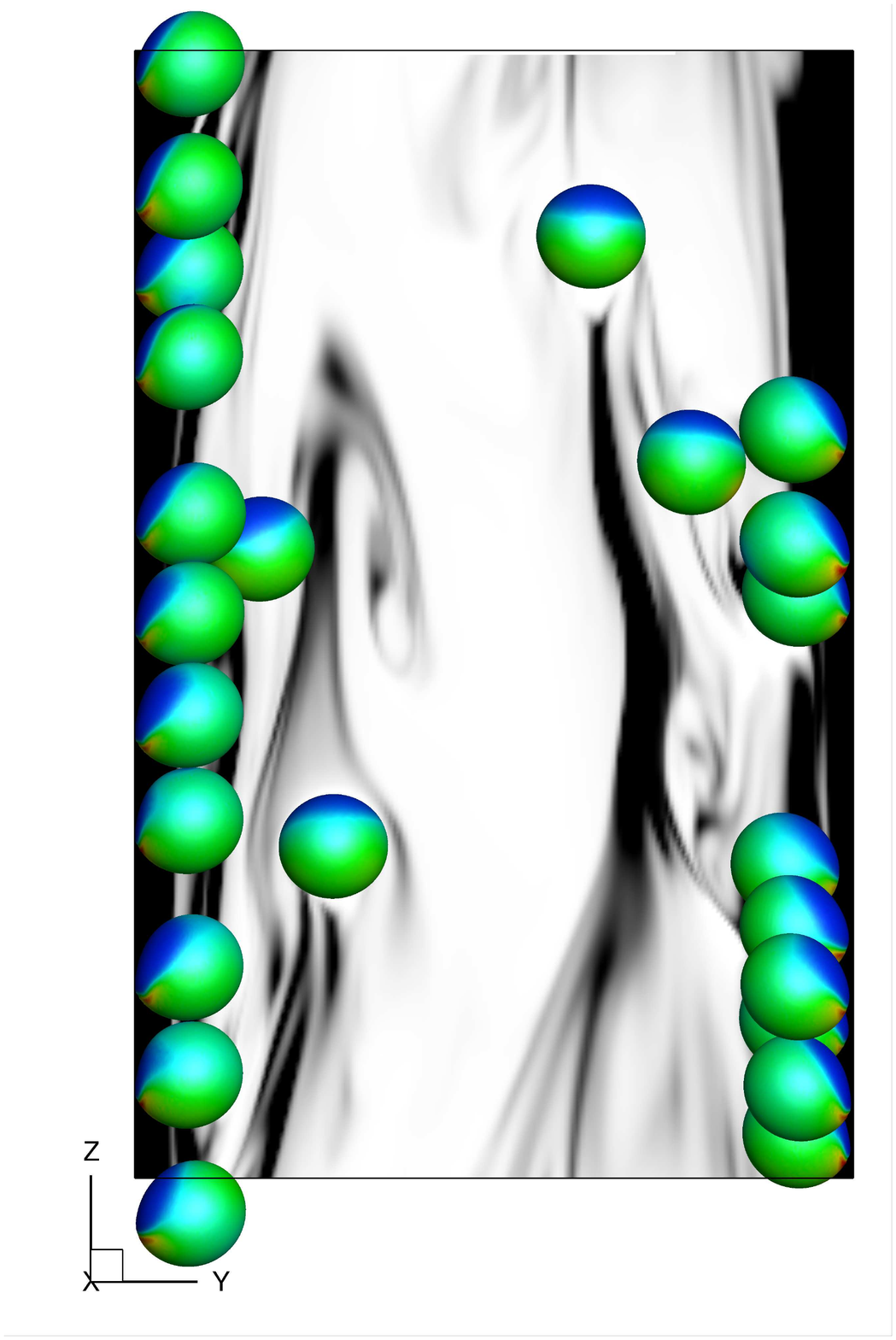}} &
  {\includegraphics[trim={0cm 0.1cm 0.3cm 0.0cm},clip,width=4cm]{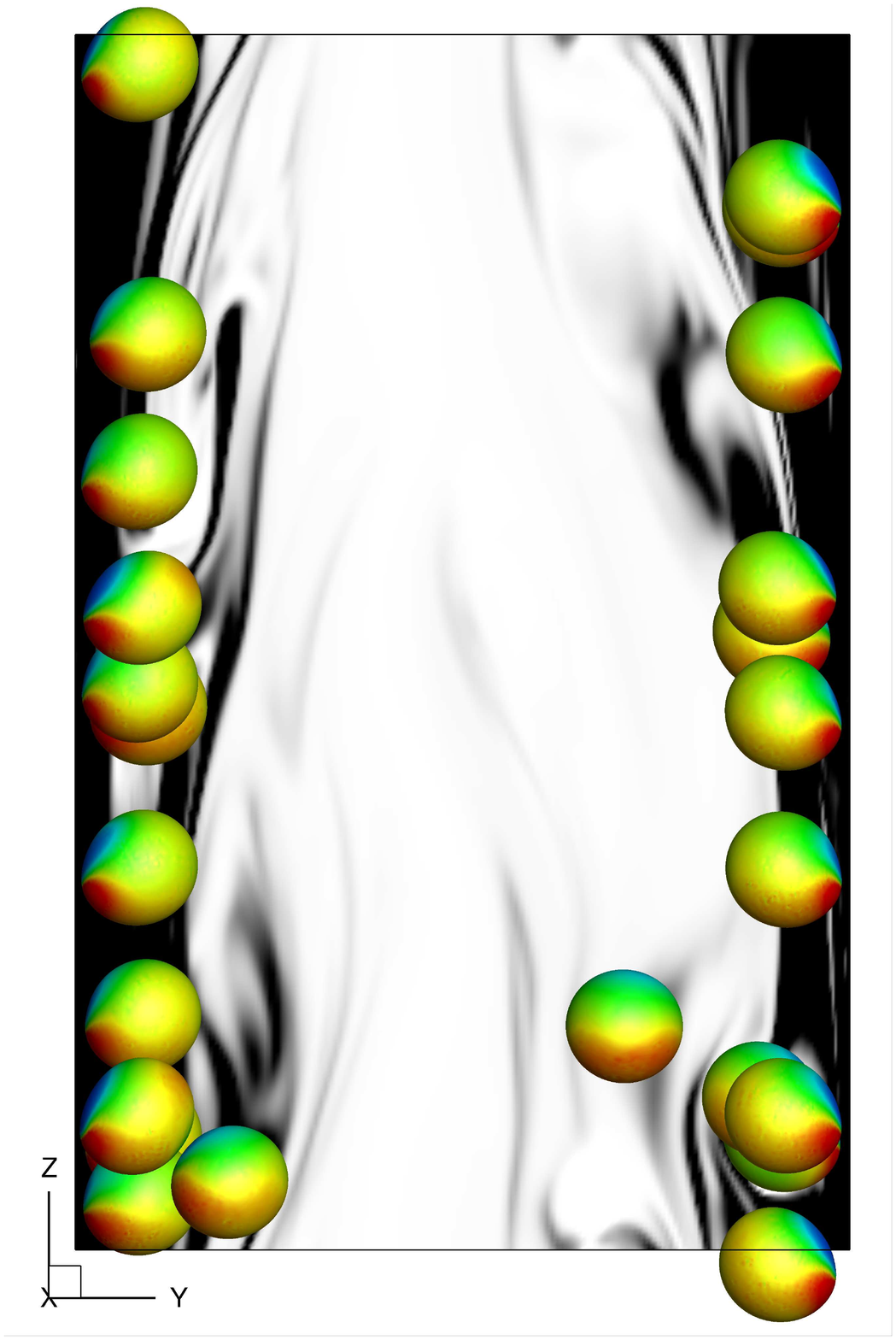}}  \\
{\includegraphics[trim={0cm 0.1cm 0.3cm 0.0cm},clip,width=4.1cm]{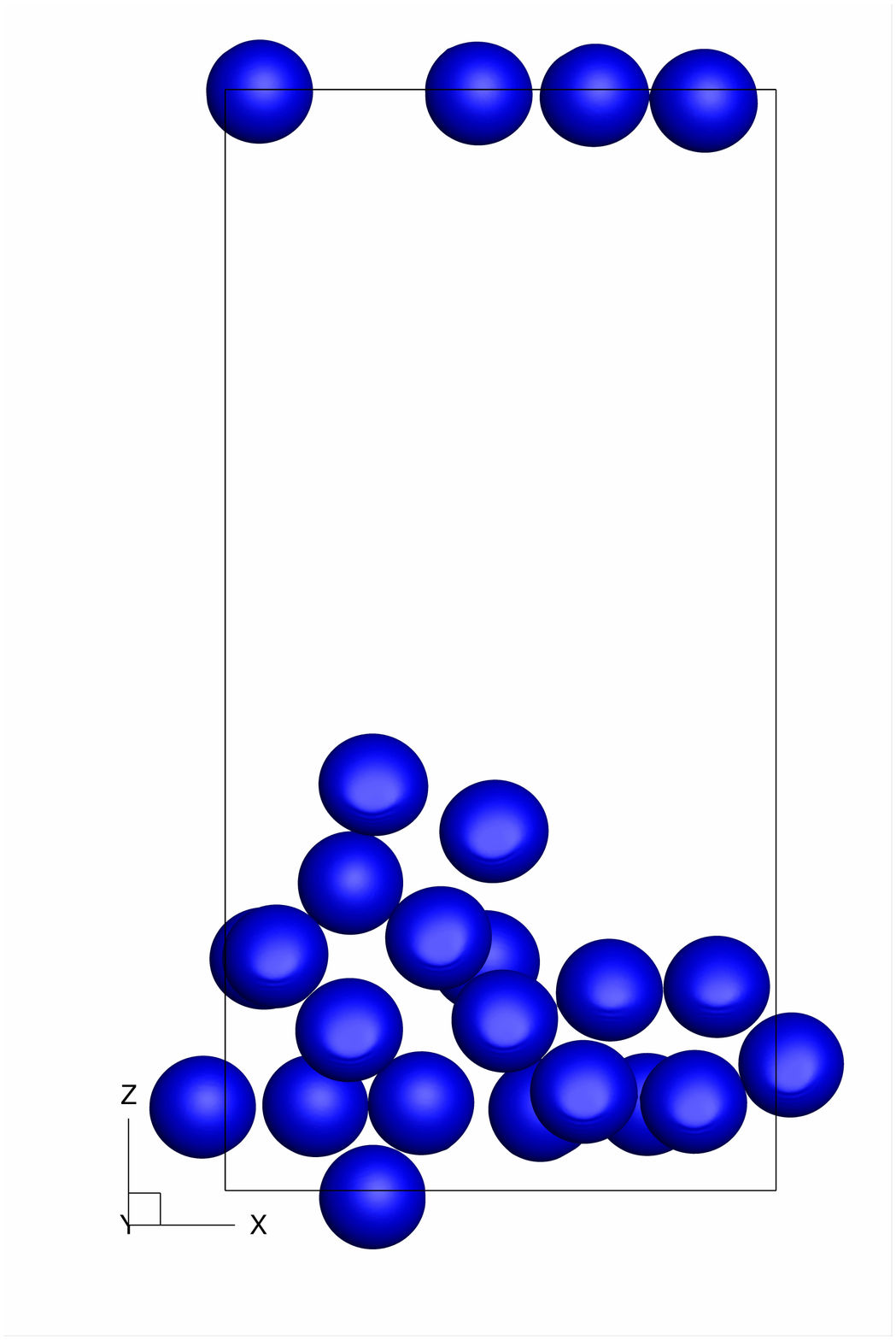}} &
{\includegraphics[trim={0cm 0.1cm 0.3cm 0.0cm},clip,width=4cm]{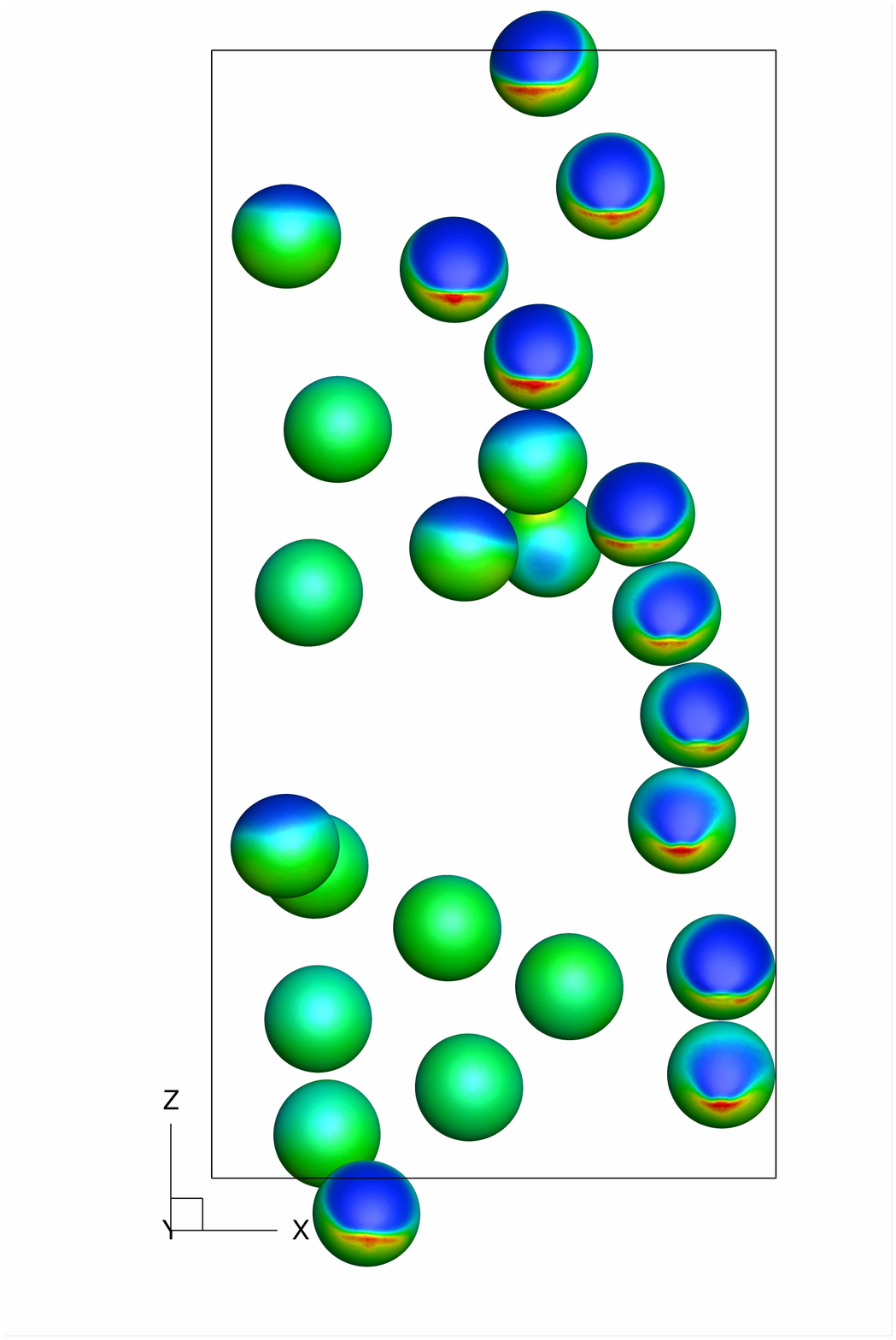}} &
{\includegraphics[trim={0cm 0.1cm 0.3cm 0.0cm},clip,width=4.3cm]{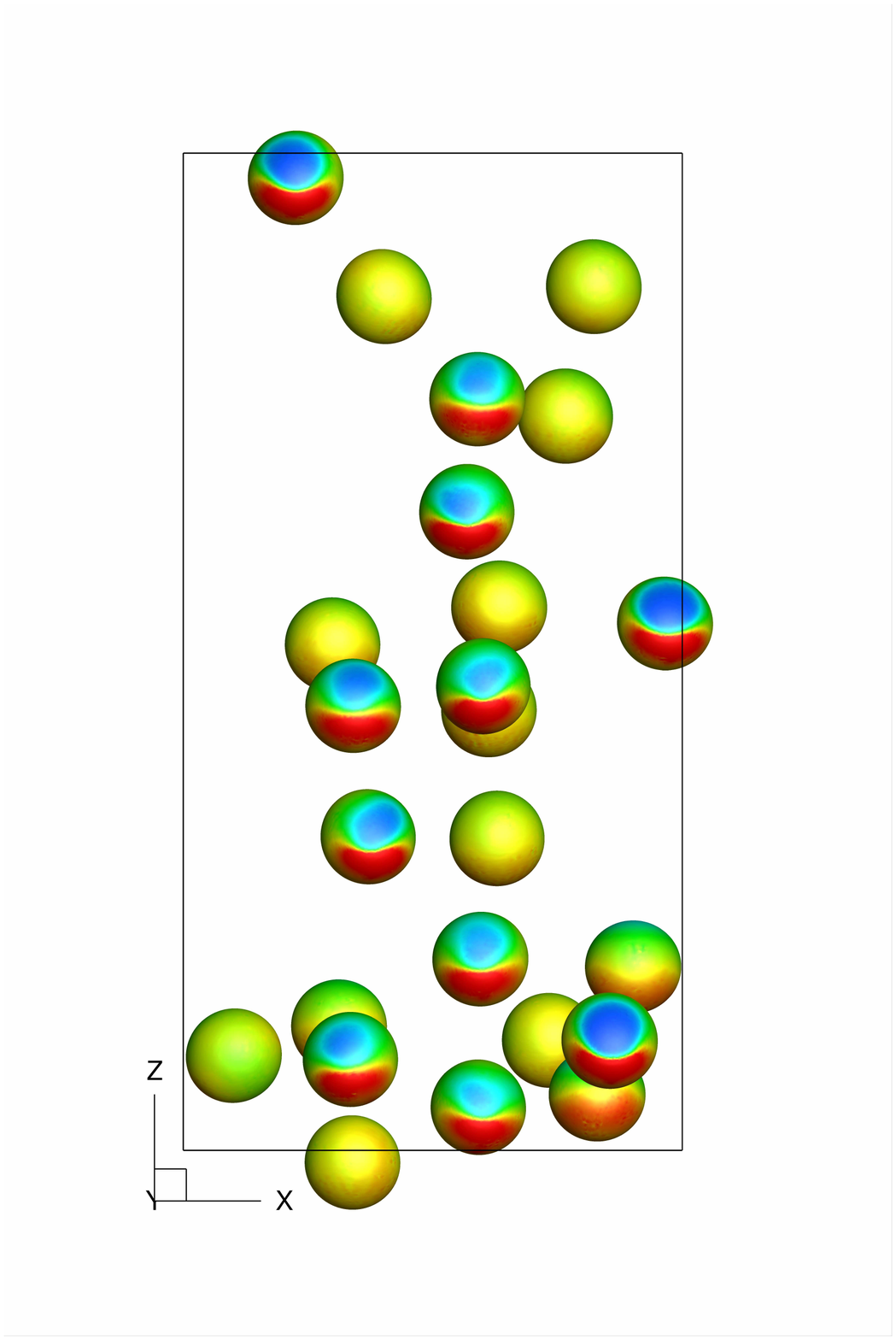}} \\                  
    $C_{\infty}=0$  &   $C_{\infty}=0.25$  & $C_{\infty}=0.5$
\end{tabular}
\end{center}
\caption{Effects of surfactant and viscoelasticity on the structure of turbulent bubbly flow. The first row shows the bubble distribution on streamwise-wall-normal plane, and the second row shows streamwise-spanwise plane. The plane contours represent normalized first normal stress difference defined as $(N_1=\frac{\tau^*_{zz}-\tau^*_{yy}}{\tau^*_{wall}})$. The contours on the bubble surface represent the interfacial surfactant concentration ($\Gamma$) with the scale ranging from $0$ (blue) to $0.15$ (red).}
\label{BubbledistV}
\end{figure}

\begin{figure}[!htb]
\setlength{\unitlength}{1cm}
\begin{center}
\begin{tabular}[c]{cc}
{\includegraphics[width=7cm]{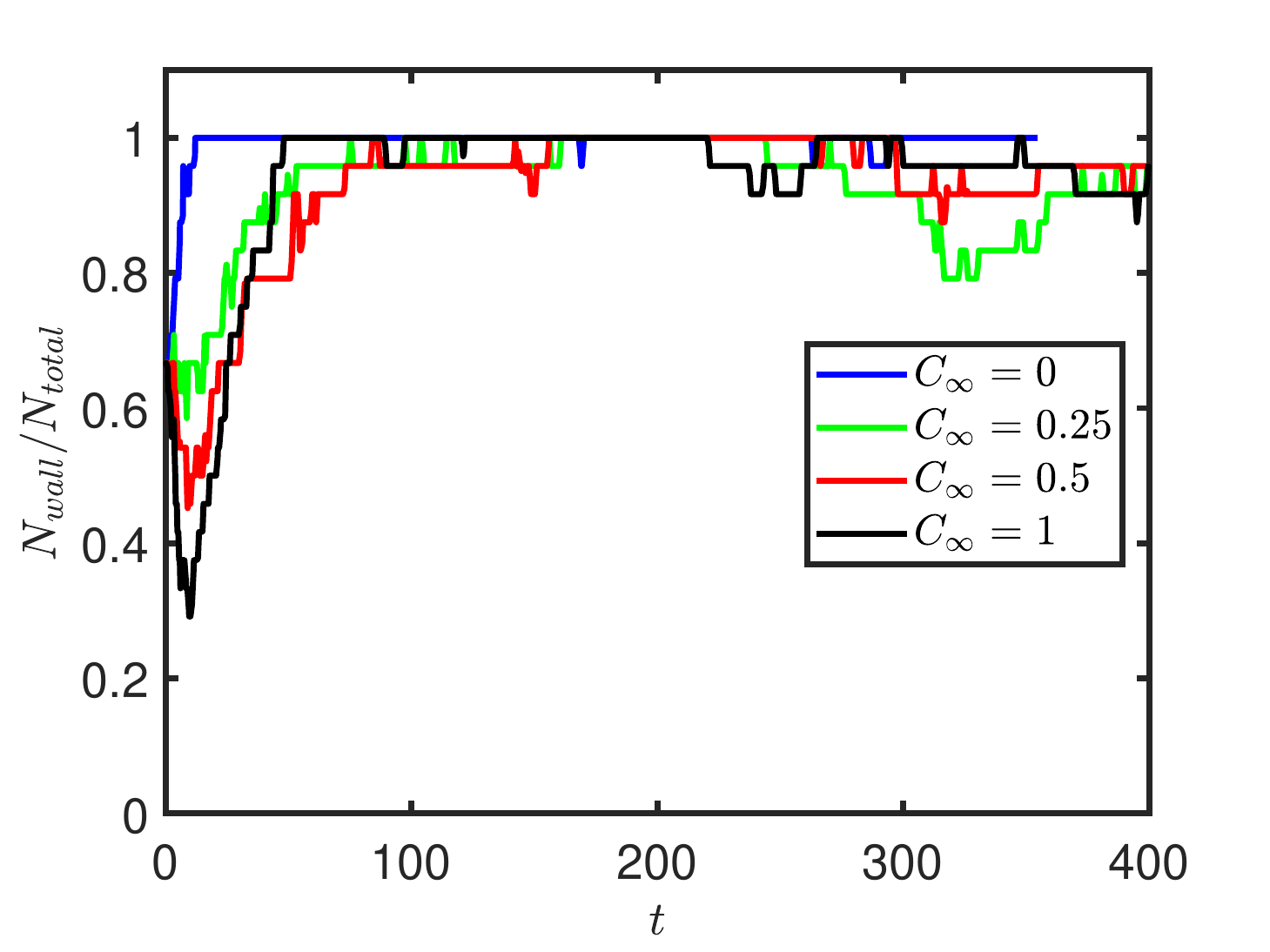}}&
{\includegraphics[width=7cm]{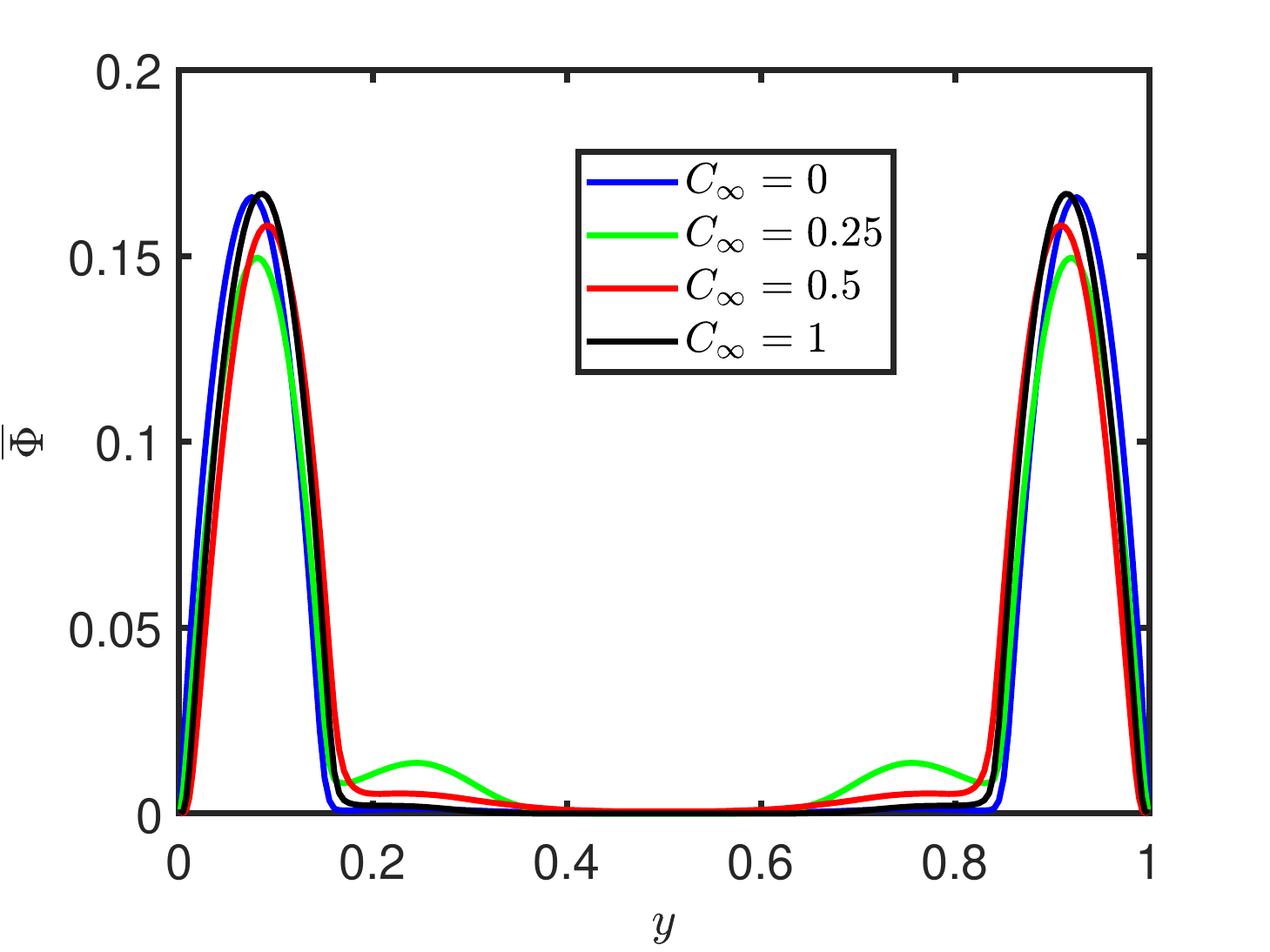}} \\
{\includegraphics[width=7cm]{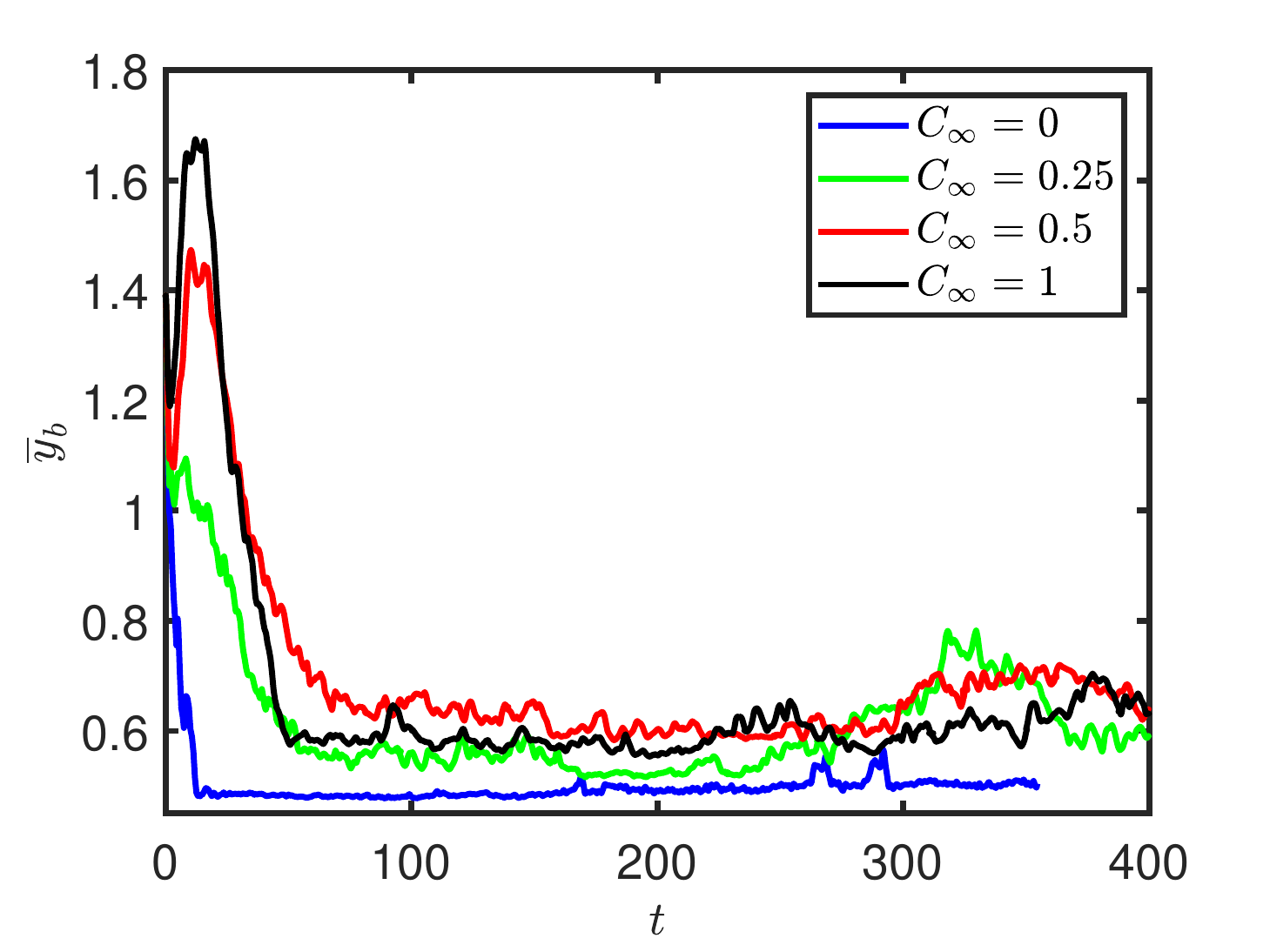}}&
{\includegraphics[width=7cm]{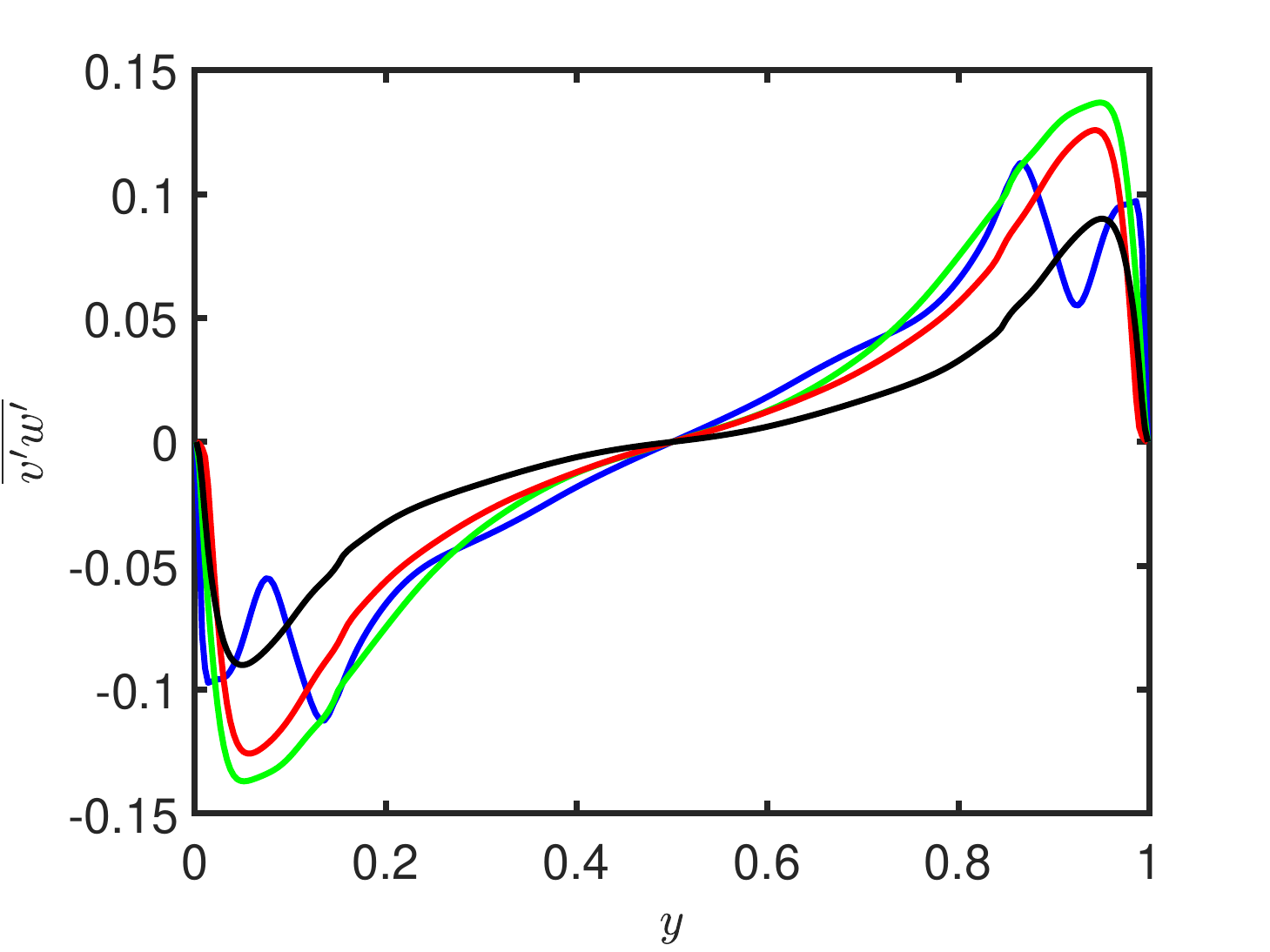}} \\
{\includegraphics[width=7cm]{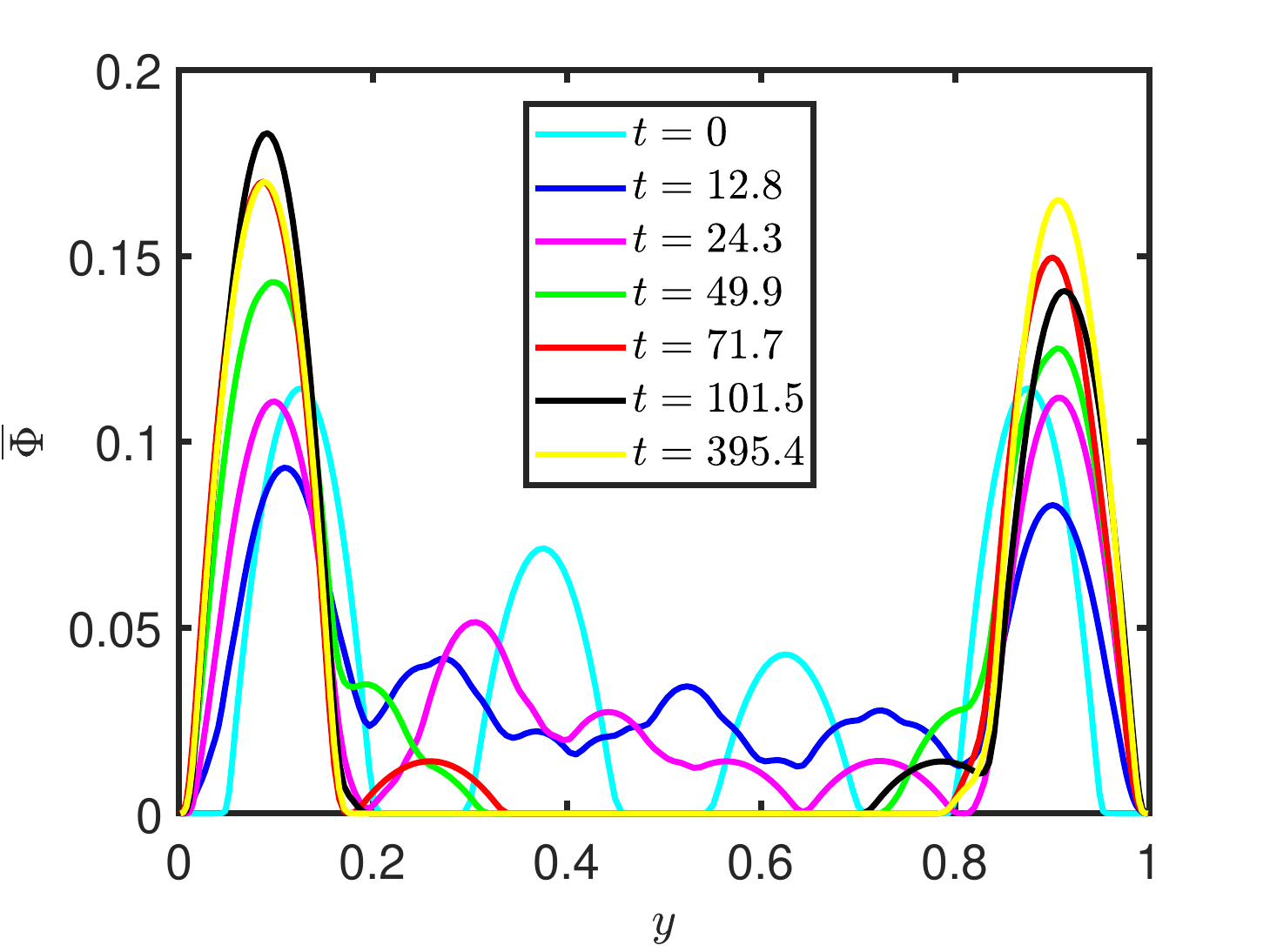}}&
{\includegraphics[width=7cm]{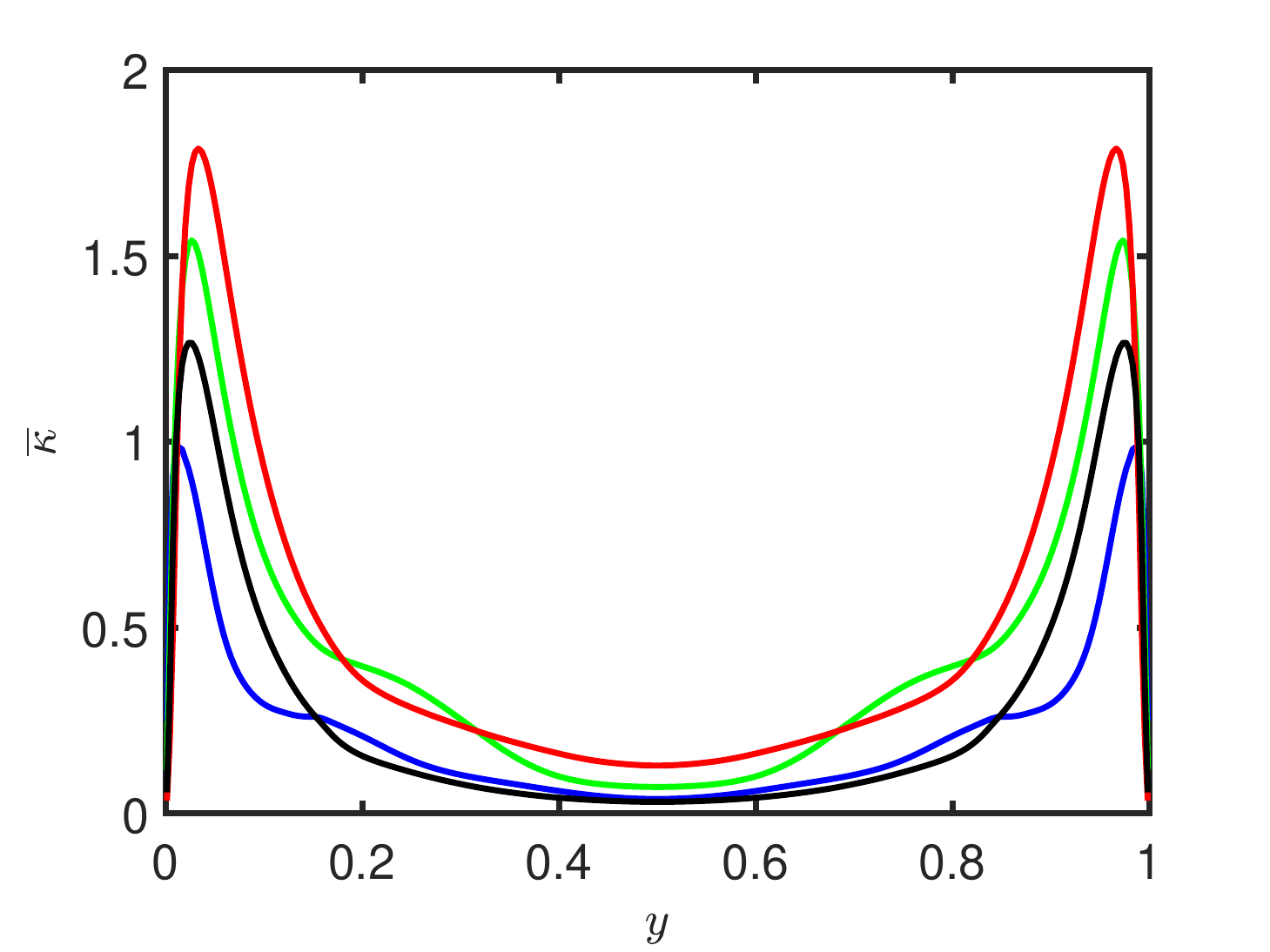}} \\
\end{tabular}
\end{center}
\vspace{-1cm}
\caption{Effects of surfactant and viscoelasticity: First column (transient results): (top) Evolution of the number of bubbles in the wall layer, (middle) the average distance of the bubbles from the wall $(\overline{y_b}=\overline{y^*_b}/d^*_b)$ and (bottom) the average void fraction for $C_\infty=0.5$. Second column (statistically stead state results): (top) average void fraction (middle) Reynolds stresses scaled by ${w^*_\tau}^2$ (bottom) turbulent kinetic energy scaled by ${w^*_\tau}^2$.}
\label{statsvisco}
\end{figure}

To examine the dynamics of turbulent bubbly flow in presence of polymers and surfactant, the simulations are performed for range of bulk surfactant concentration ($C_\infty = 0.25, 0.5, 1.0 $) while keeping the Weissenberg number ($Wi = 5$) constant. The results are shown in Fig.~\ref{flow_rateV}~\&~\ref{flow_rateV2}. As can be seen, the presence of surfactant does not show as dramatic effect as that in the case of Newtonian flow. The flow rate  decreases for all combined cases compared to the corresponding single phase flow. The amount of surfactant does not seem to be sufficient to prevent this reduction in the flow rate. The minimum reduction in the flow rate occurs for $C_\infty=0.5$ case, similar to the Newtonian flow, but the reduction in the flow rate is much higher for the combined case. To gain more insights on bubbly flow dynamics, bubble distribution is shown on streamwise-wall-normal ($z-y$) and streamwise-spanwise ($z-x$) planes in Fig.~\ref{BubbledistV}. As can be seen, for all the three combined cases, that is, clean and contaminated, bubbles accumulate at the channel walls forming bubble clusters which results in reduction of liquid flow rate. Although the surfactant concentration does not seem to be sufficient to alter the direction of migration of the bubbles, it has clearly affected the wall-layer structure. The bubbles in the horizontal clusters for the contaminated cases are not as tightly packed as observed for the clean case. Instead, they are more randomly distributed in the streamwise-spanwise ($z-x$) plane.  The contours of the first normal stress $({N_1})$ are also shown in Fig.~\ref{BubbledistV}. As can be seen, the elastic force increases with $C_\infty$ mainly due to the increase in the bulk velocity of the liquid.   

The combined effects of surfactant and viscoelasticity on the flow quantities are shown in Fig.~\ref{statsvisco}. The transient results show that the surfactant effect is prominent during the initial times until the polymers fully stretch. Thus, initially bubbles move away from the wall as the Marangoni force overcome inertial lift force. Later, the inertial and elastic lift force overcome the Marangoni forces and change the direction of migration of the bubbles towards the channel wall. Thus, wall-layers are formed for all the $C_\infty$ values unlike to the Newtonian case. The interaction between the wall-layer and core region for the $C_\infty=0$ is marginal but as $C_\infty$ increases, it can be seen that a few bubbles are moving in/out of the wall-layer continuously. As regards the Reynolds stresses, clear footprints of the bubble wall-layer can be seen near the wall. It can be also seen that the presence of surfactant has slightly improved the turbulent kinetic energy but overall it is low for all the cases due to the formation of bubble wall-layers.

\section{Conclusions}

An efficient fully parallelized finite-difference/front-tracking method has been used to examine the effects of soluble surfactant and viscoelasticity on turbulent bubbly channel flows. For the clean case, the earlier findings are verified: The clean bubbles move towards the wall due to hydrodynamic lift force and form a bubble-rich wall layer.  It is found that addition of surfactant has significant influence on the overall flow structure of turbulent bubbly channel flow. Surfactant-induced Marangoni stresses counteract the hydrodynamic lift force and push the bubbles away from the wall region. At a high enough surfactant concentration $C_\infty=0.5$, the formation of bubble wall layer is completely prevented resulting in a uniform void fraction distribution across the channel cross-section. The present results are in agreement with the earlier experimental~\citep{Takagi} and the numerical~\citep{Muradoglu, Lu} studies, i.e., the addition of surfactant prevents the formation of the wall-layers. For $C_\infty=1$, the trend reversal is also observed i,e., reforming of bubble clusters due to more uniform interfacial surfactant concentration. For the clean viscoelastic case, bubbles move toward the channel wall, similarly to the Newtonian case. The elastic force promotes the formation of the bubble wall-layer resulting in the reduction of the flow rate, which is consistent with the computational results of \citet{mukherjee2013effects}. Thus, the drag reduction effect of viscoelasticity observed for the single-phase flows diminishes due to the formation of bubble wall-layers. Finally, in the presence of combined effects of surfactant and viscoelasticity, similar wall-layer formation is observed, but the bubbles in the horizontal clusters show a more sparse distribution compared to the clean case. 

\section*{Acknowledgments}
We acknowledge financial support by the Swedish Research Council through grants No. VR2013-5789 and No. VR 2017-4809. MM acknowledges financial support from the Scientific and Technical Research Council of Turkey (TUBITAK) grant No. 115M688 and Turkish Academy of Sciences (TUBA). PC acknowledges funding from the University of Iceland Recruitment Fund grant no. 1515-151341, TURBBLY. The authors acknowledge computer time provided by SNIC (Swedish National Infrastructure for Computing).
\clearpage
\section{References}
\bibliography{AhmedAtalCF}
\end{document}